\documentclass[usenatbib]{mn2e}
\ifx\pdftexversion\undefined \usepackage{epsf,epsfig}
\else \usepackage[pdftex]{graphicx} \fi
\usepackage{lscape}
\usepackage{color,html}
\def\spose\#1{\hbox to 0pt{\#1\hss}} \def\simlt{\mathrel{\spose{\lower
3pt\hbox{$\mathchar'218$}} \raise 2.0pt\hbox{$\mathchar'13C$}}}
\def\simgt{\mathrel{\spose{\lower 3pt\hbox{$\mathchar'218$}} \raise
2.0pt\hbox{$\mathchar'13E$}}}

\author[S.J.~Oliver et al.]
{\parbox{\textwidth}{\raggedright S.J.~Oliver,$^{1}$\thanks{E-mail: \texttt{S.Oliver@Sussex.ac.uk}}
J.~Bock,$^{2,3}$
B.~Altieri,$^{4}$
A.~Amblard,$^{5}$
V.~Arumugam,$^{6}$
H.~Aussel,$^{7}$
T.~Babbedge,$^{8}$
A.~Beelen,$^{9}$
M.~B{\'e}thermin,$^{7,9}$
% ...
}\vspace{0.4cm}\\
\parbox{\textwidth}{\raggedright $^{1}$Astronomy Centre, Dept. of Physics \& Astronomy, University of Sussex, Brighton BN1 9QH, UK\\
$^{2}$California Institute of Technology, 1200 E. California Blvd., Pasadena, CA 91125, USA\\
$^{3}$Jet Propulsion Laboratory, 4800 Oak Grove Drive, Pasadena, CA 91109, USA\\
$^{4}$Herschel Science Centre, European Space Astronomy Centre, Villanueva de la Ca\~nada, 28691 Madrid, Spain\\
$^{5}$NASA, Ames Research Center, Moffett Field, CA 94035, USA\\
$^{6}$Institute for Astronomy, University of Edinburgh, Royal Observatory, Blackford Hill, Edinburgh EH9 3HJ, UK\\
$^{7}$Laboratoire AIM-Paris-Saclay, CEA/DSM/Irfu - CNRS - Universit\'e Paris Diderot, CE-Saclay, pt courrier 131, F-91191 Gif-sur-Yvette, France\\
$^{8}$Astrophysics Group, Imperial College London, Blackett Laboratory, Prince Consort Road, London SW7 2AZ, UK\\
$^{9}$Institut d'Astrophysique Spatiale (IAS), b\^atiment 121, Universit\'e Paris-Sud 11 and CNRS (UMR 8617), 91405 Orsay, France\\
% ...
$^{13}$Departamento de Astrof{\'\i}sica, Universidad de La Laguna (ULL), E-38205 La Laguna, Tenerife, Spain\\
% ...
}}

\title[HerMES]{The {\em Herschel}\footnote{{\em Herschel} is an ESA space observatory with science instruments provided
by European-led Principal Investigator consortia and with important participation from NASA.} Multi-tiered Extragalactic Survey: HerMES} \date{Dec, 21th 2010}

\begin{document}

\maketitle


\begin{abstract} 
The {\em Herschel} Multi-tiered Extragalactic Survey, HerMES, is a legacy program designed to map a set of nested fields totalling {\color{black} $\sim 380~{\rm deg}^2$}.  Fields range in size from 0.01 to $\sim20~{\rm deg}^2$, using {\em Herschel}-SPIRE (at 250, 350 and 500$\,$\micron), and {\em Herschel}-PACS (at 100 and 160$\,$\micron), with an additional wider component of $270\, {\rm deg}^2$ with SPIRE alone. 
These bands cover the peak of the redshifted thermal spectral energy distribution from interstellar dust and thus capture the re-processed optical and ultra-violet radiation from star formation that has been absorbed by dust, and are critical for forming a complete multi-wavelength understanding of galaxy formation and evolution.  

The survey will detect of order 100,000 galaxies at $5\sigma$ in some of the best studied fields in the sky.  Additionally, HerMES is closely coordinated with the PACS Evolutionary Probe survey.  Making maximum use of the full spectrum of ancillary data, from radio to X-ray wavelengths, {\color{black} it is designed} to: facilitate redshift determination; rapidly identify unusual objects; and understand the relationships between thermal emission from dust and other processes.  Scientific questions HerMES will be used to answer include: the total infrared emission of galaxies;  the evolution of the luminosity function; the clustering properties of dusty galaxies; and the properties of populations of galaxies which lie below the confusion limit through lensing and statistical techniques.  

This paper defines the survey observations and data products, outlines the primary scientific goals of the HerMES team,  and reviews some of the early results.

\end{abstract}
\begin{keywords}
surveys -- infrared: galaxies -- submillimetre: galaxies -- galaxies: evolution
\end{keywords}

\section{Introduction \& Science Goals}


Understanding how galaxies form and evolve over cosmological time is a key goal in astrophysics.   
Over the last decade our understanding of the background cosmology has improved to such an extent \citep[e.g.][]{Spergel:2003} that we think we have a reasonable understanding of the formation of structure in the 
underlying dark matter distribution \citep[e.g.][]{Springel:2006}. 
However, galaxy formation and evolution are driven by dissipative, non-linear processes within the potential wells of virialized dark matter halos which are much more complex physical processes which have defied full modelling.  Observations play a critical role in constraining models of galaxy formation, the evolution of star-formation
activity, and the various roles played by galaxy stellar mass, dark matter halo mass, and environment.  


The central importance of far-infrared (FIR) and sub-millimetre (sub-mm) observations becomes clear when one realizes that the approximately half of all the luminous power \citep{puget96,Fixsen1998,Lagache1999} which makes up the extra-galactic background radiation -- power which originated from stars and active galactic nuclei (AGN) -- was emitted at optical/ultraviolet wavelengths, absorbed by dust, and reradiated in the FIR/sub-mm.  To form a complete picture of the evolution of galaxies, the optical regime alone cannot be used to fully trace the activity \citep[e.g., the brightest sub-mm galaxy in the Hubble Deep Field is not even detected in the optical][]{Dunlop:2004}.  
Furthermore, sub-mm observations provide a bridge in both wavelength and redshift  between the  $z > 2$ Universe, primarily probed on the Rayleigh-Jeans side of the spectral energy distribution (SED) by ground based sub-mm telescopes, and the lower-$z$ Universe, sampled on the Wein side of the SED by {\em Spitzer}. 

FIR/sub-mm luminosity is thought to arise primarily from dust heated by the massive stars in star formation regions and so may be used as a direct estimate of star formation activity.  
Additional contributions are expected to arise from dusty tori surrounding AGN at shorter wavelengths, and there may be non-negligible contributions from the illumination of dust by evolved stars. 

\begin{figure}
\includegraphics[width=8cm]{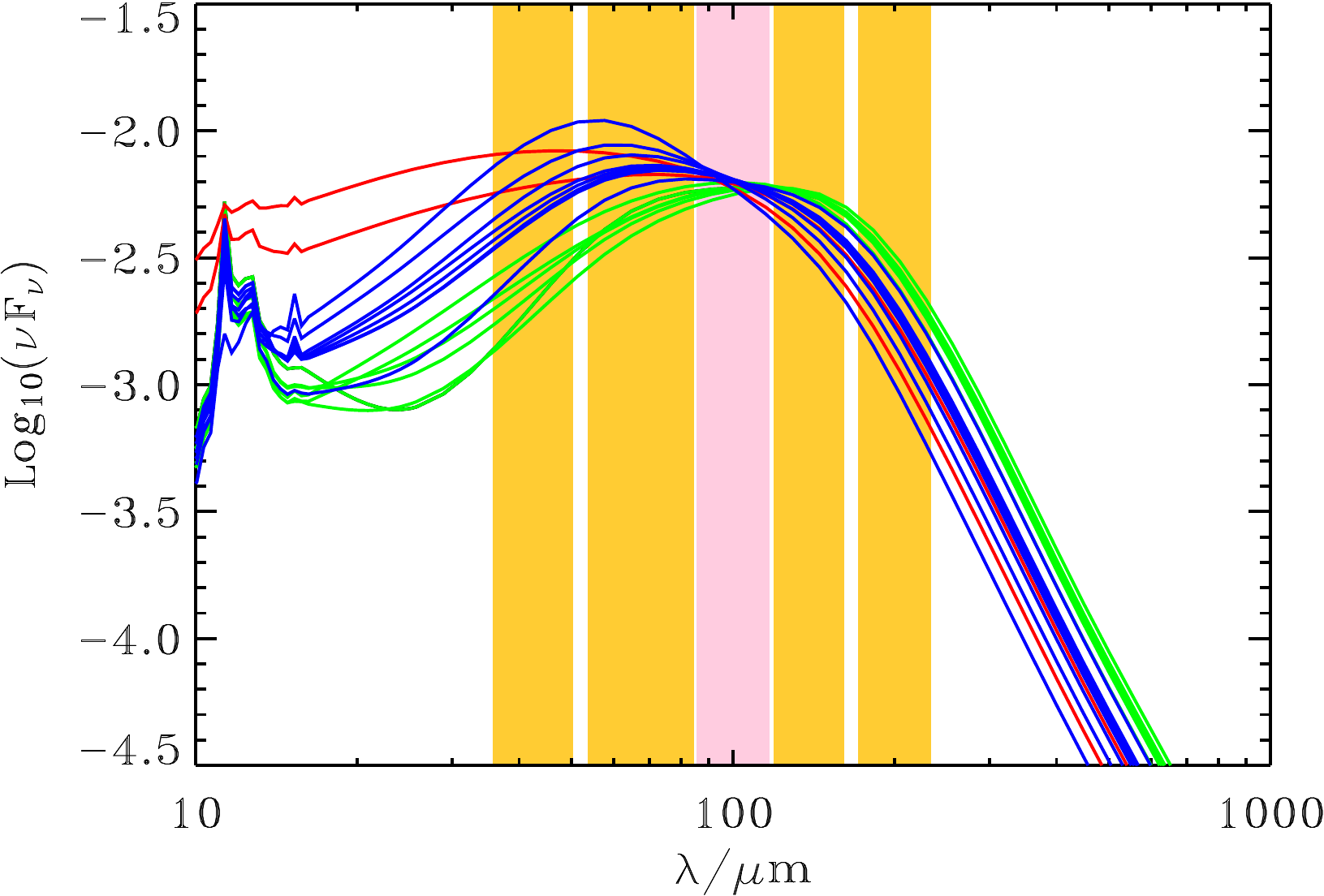}
\caption{Model {\color{black} spiral (green), star-burst  galaxy (blue) and AGN (red)} Spectral Energy Distributions (SEDs) normalised to the same $L_{\rm FIR}$ and plotted in their rest-frame with the {\em Herschel}-PACS and {\em Herschel}-SPIRE bands at  {\color{black} $\lambda=100, 160, 250, 350$ and $500\,$\micron\ plotted at $\lambda/(1+z)$} for a  galaxy at $z=1.5$.  Note that the {\em Herschel}-SPIRE band at 250$\,$\micron measures a similar flux density for all and so is a reasonable proxy for the $L_{\rm FIR}$ for these templates.}\label{fig:sed}
\end{figure}

Previous surveys from space-based observatories: {\em IRAS} \cite[e.g.][]{Saunders90, Oliver92}; {\em ISO} \citep[e.g.][and references therein]{Elbaz1999, Oliver2002};  and {\em Spitzer} \citep[e.g.][and references therein]{Shupe2008, Frayer2009};  and at sub-mm wavelengths from the ground with:
SCUBA at 850$\,$\micron\ \citep[e.g.][]{Eales:1999,Hughes:1998,Smail:1997, Coppin2006}, Bolocam \citep[e.g.]{Maloney2005}; 
SHARCII \citep[e.g.][]{Khan2007}; 
MAMBO \citep[e.g.][]{Greve2008};
LABOCA \citep[e.g.][]{Weiss2009}; and
AzTEC \citep[e.g.][]{Scott2010}, 
demonstrated strong evolution in galaxies at both mid-infrared (MIR) and FIR wavelengths. 
This evolution is attributed to a decline in the average star-formation density with time, and particularly a decline in the 
role of the more luminous infrared galaxies (LIRGs), which are thought to be the progenitors of massive galaxies today (e.g. \citealt{Le-Floch:2005}).  

This strong evolution has been challenging for physical models of galaxy formation to reproduce.  They find they must invoke drastic modifications, such as altering the initial mass function (e.g. \citealt{Baugh2005}), in order to match these observations as well as optical and near infrared constraints on the stellar mass today.  

\begin{figure*}
\includegraphics[width=8cm]{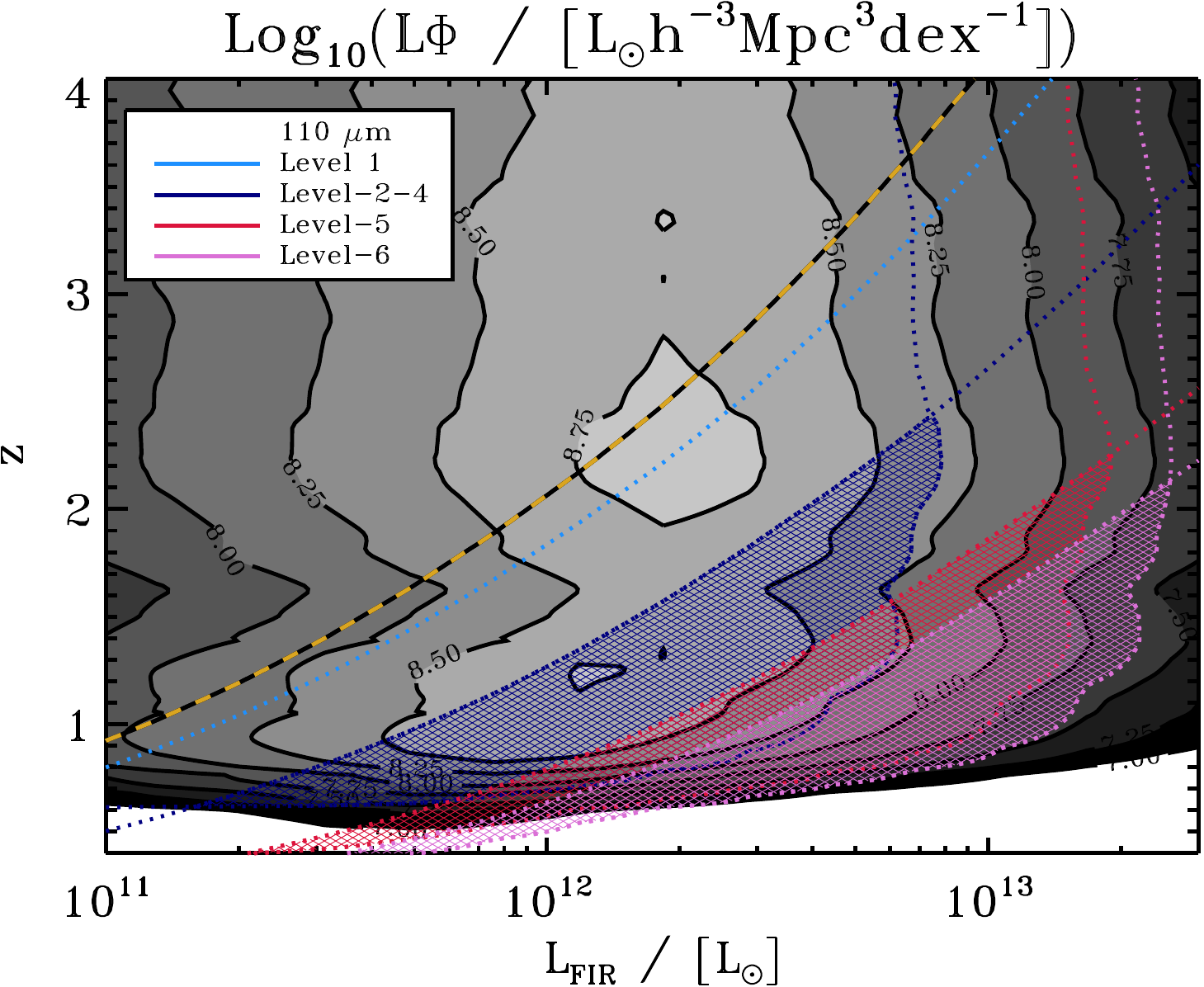}
\includegraphics[width=8cm]{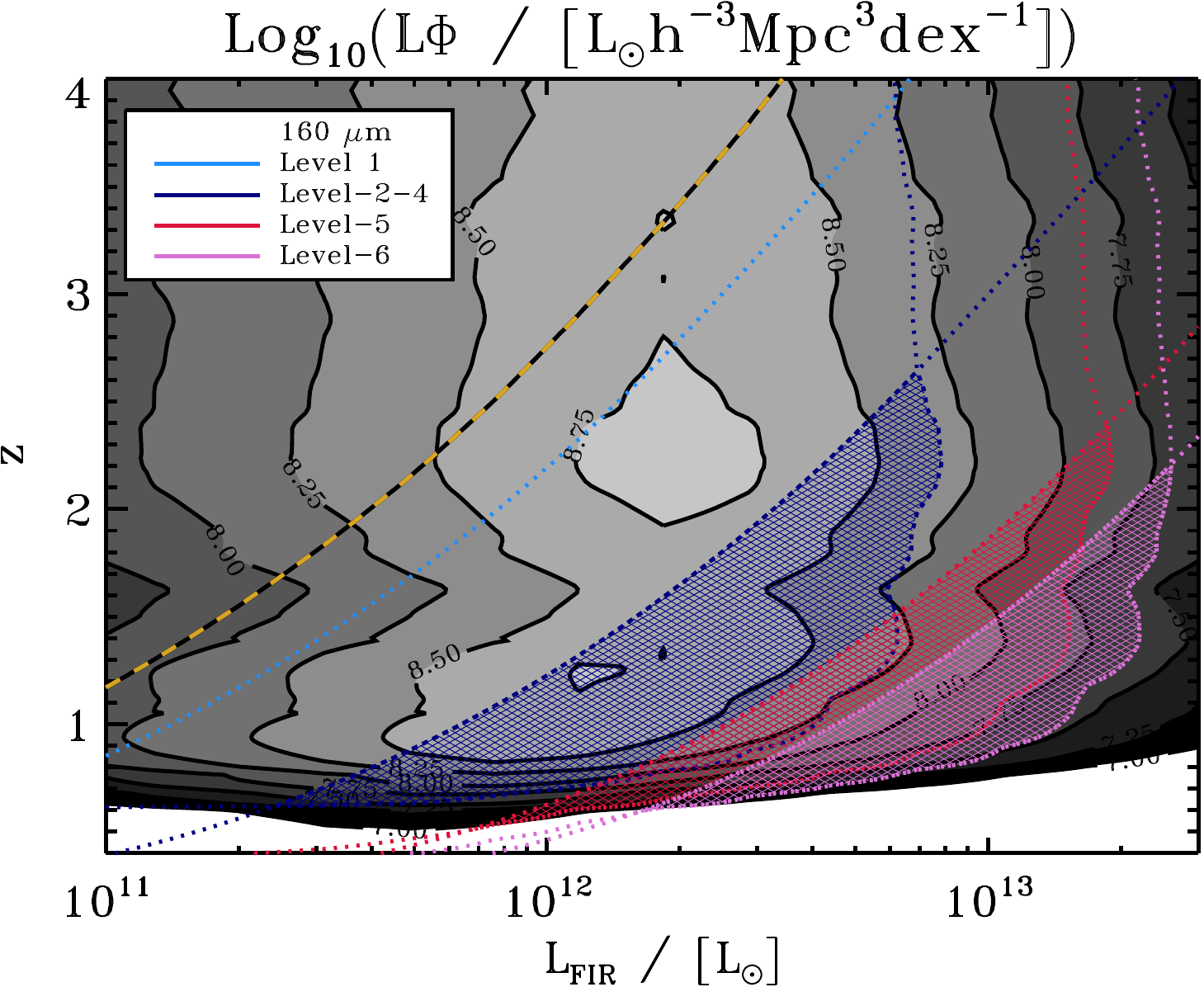}
\includegraphics[width=8cm]{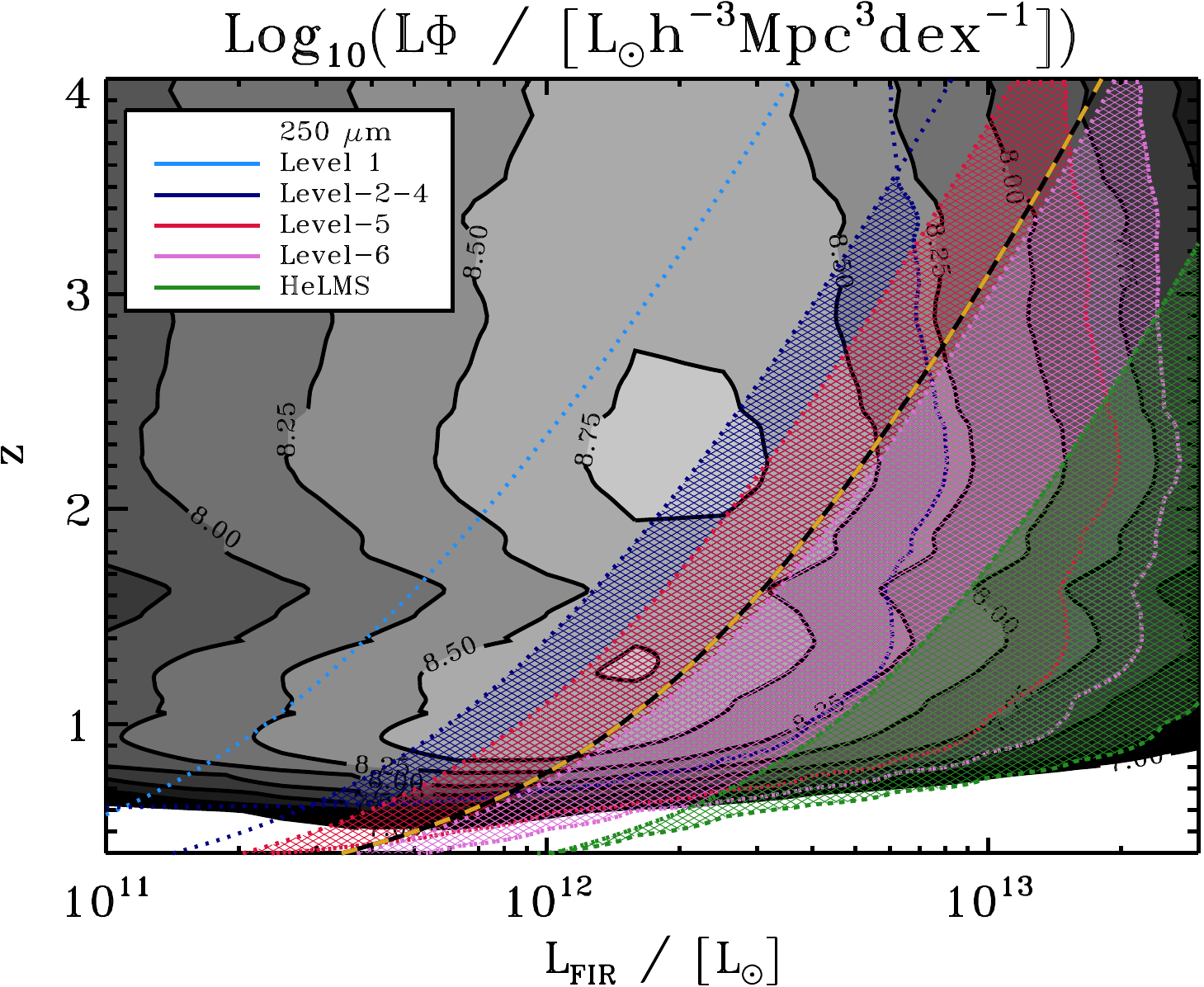}
\includegraphics[width=8cm]{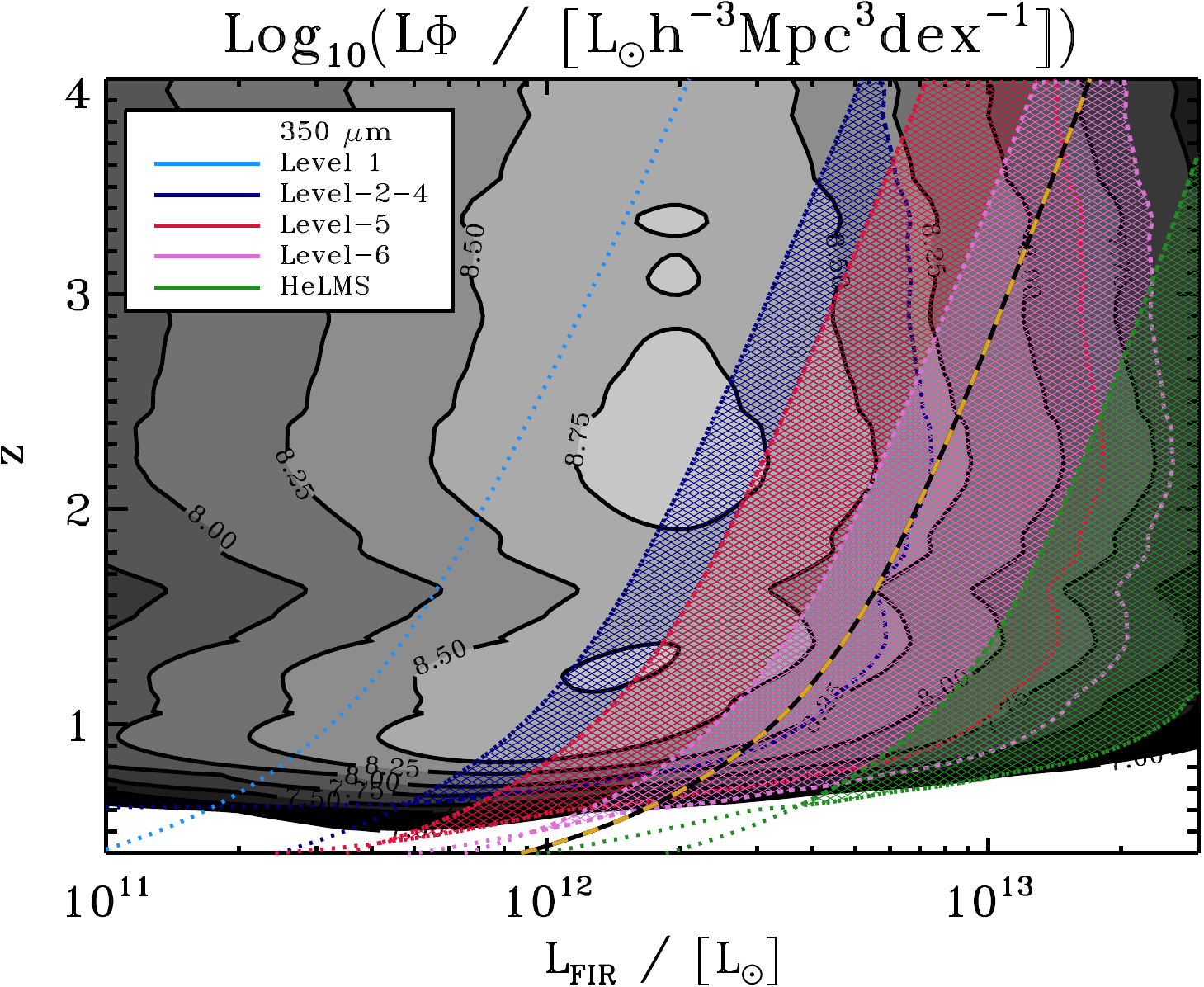}
\includegraphics[width=8cm]{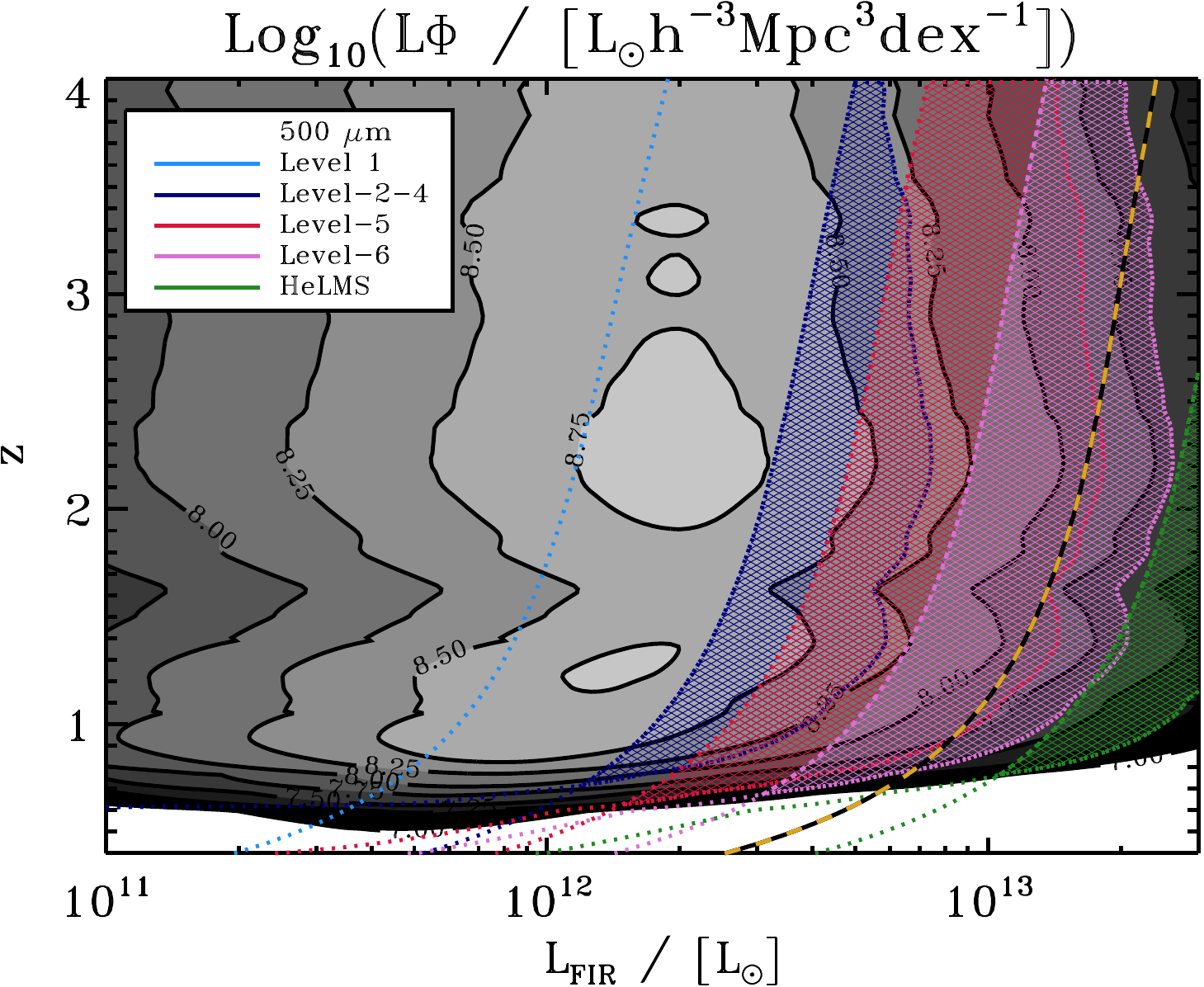}
\caption{Far infrared luminosity density in $\log_{10} ({\rm L}_{\sun} h^{-3}{\rm Mpc}^3{\rm dex}^{-1})$ (grey-scale and contour diagram) as a function of far infrared luminosity ($x$-axis) and redshift ($y$-axies) -- from the model of \protect\citealt{Lagache:2003}.  The power of different survey elements to probe this space are indicated by overlays.  Each panel shows survey elements at different wavelengths; reading left-to-right from the top they are 100, 160, 250, 350 and 500$\,$\micron.  Surveys are deemed to properly sample the space if they can detect galaxies of these FIR luminosities at the 5-$\sigma$ instrumental noise level and with more than 75 galaxies in bins of $\Delta\log L\,\Delta z=0.1$. These two constraints are marked with dotted lines and are hatched. The different survey levels defined in Table~\protect\ref{tab:250counts} are shown with: Levels~2--4 -- blue; Level~5  -- red; Level~6 -- magenta and HeLMS- -- green. Level-1 (cyan) does not have enough volume to satisfy the number of galaxies criterion and so only the instrumental noise limit is shown. The 5$\sigma$ confusion noise levels (after 5$\sigma$ clipping) from \protect\citet[at 100 and 160\micron]{Berta2011} and \protect\citet[at 250, 350 and 500\micron]{Nguyen2010}  with yellow/black lines.  Note the bimodal peaks at $z \sim 1$ and $z \sim 2.5$} \label{fig:lumden}
\end{figure*}

Using a different approach, phenomenological galaxy population models attempt to describe what is currently observed and also predict what we would expect for {\em Herschel}.  
Different groups use different combinations of galaxy populations to reproduce the observations; for example, \citet[and Fig.~\ref{fig:lumden}]{Lagache:2003} use two peaks of luminosity density at $z\sim 1 $ and $z\sim2$ to describe the data, which are not seen in other models. Such differences between the pre-{\em Herschel} models indicate the lack of constraint on the spectral energy distributions and redshift distributions. 



The potential of sub-mm surveys has been demonstrated by the BLAST telescope \citep{blast}. BLAST was a balloon-borne telescope with a focal plane instrument based on the SPIRE \citep{Griffin2010}  photometer design and using similar detectors tailored to higher photon loading, and was a successful technical and scientific pathfinder for
SPIRE on {\em Herschel}, probing the wavelength regime where the SED of redshifted galaxies and the infrared background peak.
  

The {\em Herschel Space Observatory} \citep{Pilbratt2010} is carrying out surveys of unprecedented size and depth, vastly improving the state of observations in this under-explored waveband.  
The imaging instruments SPIRE \citep{Griffin2010} and PACS \citep{Poglitsch2010}, which together fully constrain the peak of the FIR/sub-mm
background, allow us to thoroughly investigate the sources in the infrared background and characterize their total obscured emission (see e.g. Fig.~\ref{fig:sed}).  

The {\em Herschel} Multi-tiered Extra-galactic Survey (HerMES\footnote{http://hermes.sussex.ac.uk. Hermes is also the Olympian messenger god, ruler of travellers, boundaries, weights and measures.}) is a Guareenteed Time Key Program on {\em Herschel} which will provide a legacy survey of star forming galaxies over the wavelengths at which the galaxies and infrared background peak.  
The majority of science goals require multi-wavelength support and the fields we have chosen are among the best in the sky for multi-wavelength coverage (see Section~\ref{sec:ancillary}) maximising their legacy value. 

In Section~\ref{sec:design} we define the survey. In Section~\ref{sec:method} we described some of our goals and early results.  In Section~\ref{Sec:Data} we outline our expected data products and delivery time-scales before concluding in Section~\ref{sec:conc}.



\section{Survey Design}\label{sec:design}

Our survey is defined by Astronomical Observing Requests (AORs).
For convenience we have grouped the AORs by sets, which in this paper are identified with numbers, e.g., ELAIS N2 SWIRE is \#41. A summary of the AOR sets is given in Table~\ref{tab:AORs}. Details of the observing modes can be found in the {\em Herschel} observers' manuals {\color{black}(available from \verb+http://herschel.esac.esa.int/Documentation.shtml+)}. 

 Detector hits maps{\color{black}\footnote{\color{black} These maps and Table~\ref{tab:AORs} gives coverage for SPIRE observations as counts of 250 \micron\ detector samples per 6\arcsec$\times$6\arcsec\ pixel. This can be converted to a bolometer ``exposure'' time per pixel by dividing by the sampling frequencies of 18.6$\,$ Hz for SPIRE scanning at nominal and fast rates and 10$\,$ Hz for parallel mode.  The hits in other arrays can be estimated by scaling by the numbers of detectors in the arrays (129, 88, 43) and the pixel sizes (6\arcsec, 10\arcsec, 12\arcsec).}}, which accurately define the coverage of the survey and should be used for any detailed planning of complementary surveys, are provided on our web site \verb+http://hermes.sussex.ac.uk+.  We also provide files which define the approximate boundaries of homogenous regions (e.g. as marked in Fig.~\ref{fig:covmono}). These survey definition products are updated as the survey progresses. Our sensitivities have been quoted using official mission values given in Table~\ref{tab:sense}. 

The current AORs which define our program can be retrieved  from the {\em Herschel} Science Archive \protect\verb+http://herschel.esac.esa.int/Science_Archive.shtml+ using HSpot and the proposal IDs SDP\_soliver\_3 and KPGT\_soliver\_1 and GT2\_mviero\_1.

Here we summarise the basis of our survey design.

\begin{table*}
\begin{tabular}{rl |l |l |r |r |r |r |r |r |r |r |r |r |r |l | r r r }
\hline
  \multicolumn{1}{|c|}{Set} &
  \multicolumn{1}{c|}{Level} &
  \multicolumn{1}{c|}{Target} &
  \multicolumn{1}{c|}{Mode} &
  \multicolumn{1}{c|}{$N_{\rm AOR}$} &
  \multicolumn{1}{c|}{$T$} &
  \multicolumn{1}{c|}{$N_{\rm rep}$} &
  \multicolumn{1}{c|}{$N_{\rm samp}$} &
  \multicolumn{1}{c|}{$l_1$}&
  \multicolumn{1}{c|}{$l_2$} &
  \multicolumn{1}{c|}{$\theta$} &
  \multicolumn{1}{c|}{$\Omega_{\rm nom}$} &
  \multicolumn{1}{c|}{$\Omega_{\rm max}$} &
  \multicolumn{1}{c|}{$\Omega_{\rm good}$} &
  \multicolumn{1}{c|}{DR} \\
    \multicolumn{1}{|c|}{}&
  \multicolumn{1}{c|}{}&
  \multicolumn{1}{c|}{}&
  \multicolumn{1}{c|}{}&
  \multicolumn{1}{c|}{}&
  \multicolumn{1}{c|}{[hr]}&
  \multicolumn{1}{c|}{}&
  \multicolumn{1}{c|}{}&
  \multicolumn{1}{c|}{[\arcmin]} &
  \multicolumn{1}{c|}{[\arcmin]} & 
  \multicolumn{1}{c|}{[deg]}& 
  \multicolumn{1}{c|}{[deg$^2$]}&
  \multicolumn{1}{c|}{[deg$^2$]}&
  \multicolumn{1}{c|}{[deg$^2$]}&
  \multicolumn{1}{c|}{}
  \\
\hline
  1 & CD & Abell 2218 & Sp. Nom. & 2 & 9.29 & 100 & 1118 & 4 & 4 & 84 &  & 0.14 & 0.10 & SDP\\
  2 & CD & Abell 1689 & Sp. Nom. & 8 & 1.97 & 48 & 235 & 4 & 4 & 18 &  & 0.11 & 0.08 & \\
  3 & CD & MS0451.6-0305 & Sp. Nom. & 8 & 1.97 & 48 & 235 & 4 & 4 & 5 &  & 0.11 & 0.08 & DR1\\
  4 & CS & RXJ13475-1145 & Sp. Nom. & 8 & 1.97 & 48 & 234 & 4 & 4 & 17 &  & 0.11 & 0.08 & \\
  5 & CS & Abell 1835 & Sp. Nom. & 8 & 1.97 & 48 & 236 & 4 & 4 & 16 &  & 0.11 & 0.08 & \\
  6 & CS & Abell 2390 & Sp. Nom. & 8 & 1.97 & 48 & 235 & 4 & 4 & 81 &  & 0.11 & 0.08 & \\
  7 & CS & Abell 2219 & Sp. Nom. & 8 & 1.97 & 48 & 234 & 4 & 4 & 66 &  & 0.11 & 0.08 & DR1\\
  8 & CS & Abell 370 & Sp. Nom. & 8 & 1.97 & 48 & 233 & 4 & 4 & 70 &  & 0.11 & 0.08 & \\
  9 & CS & MS1358+62 & Sp. Nom. & 8 & 1.97 & 48 & 235 & 4 & 4 & 76 &  & 0.11 & 0.08 & \\
  10 & CS & Cl0024+16 & Sp. Nom. & 8 & 1.97 & 48 & 235 & 4 & 4 & 61 &  & 0.11 & 0.08 & \\
  11 & CH & MS1054.4-0321 & Sp. Nom. & 8 & 2.18 & 16 & 131 & 15 & 10 & 22 &  & 0.24 & 0.16 & \\
  12 & CH & RXJ0152.7-1357 & Sp. Nom. & 8 & 2.18 & 16 & 131 & 15 & 10 & 165 &  & 0.24 & 0.16 & \\
  13 & L1 & GOODS-S & Sp. Nom. & 76 & 20.22 & 76 & 730 & 20 & 20 & 14 &  & 0.51 & 0.35 & \\
  22 & L2 & COSMOS & Sp. Nom. & {\color{black} 24} & 50.13 & {\color{black} 24} & {\color{black} 388} & 85 & 85 & 70 &  & {\color{black} 3.49} & {\color{black} 2.82} & \\
  14 & L2 & GOODS-N & Sp. Nom. & 1 & 13.51 & 30 & 416 & 30 & 30 & 42 &  & 0.64 & 0.55 & SDP\\
  15 & L2 & ECDFS & Sp. Nom. & 19 & 8.78 & 19 & 232 & 30 & 30 & 44 &  & 0.79 & 0.58 & DR1\\
  17 & L3 & Groth Strip & Sp. Nom. & 7 & 3.54 & 7 & 85 & 67 & 10 & 130 &  & 0.82 & 0.60 & DR1\\
  18 & L3 & Lockman-East ROSAT & Sp. Nom. & 7 & 3.2 & 7 & 87 & 30 & 30 & 77 &  & 0.77 & 0.57 & \\
  18B & L3 & Lockman-East Spitzer & Sp. Nom. & {\color{black} 4} & {\color{black} 4.53} & {\color{black}4} &  {\color{black} 32} & {\color{black} 80}  & {\color{black} 40} & {\color{black} 149} &  & {\color{black} 1.78} & {\color{black} 1.40} & \\
  19 & L3 & Lockman-North & Sp. Nom. & 1 & 3.91 & 7 & 104 & 35 & 35 & 1 &  & 0.74 & 0.65 & SDP\\
  23 & L4 & UDS & Sp. Nom. & 7 & 10.54 & 7 & 110 & 66 & 66 & 20 &  & 2.46 & 2.02 & \\
  24 & L4 & VVDS & Sp. Nom. & 7 & 10.39 & 7 & 110 & 66 & 66 & 21 &  & 2.46 & 2.02 & \\
 22B & L5 & COSMOS HerMES &  Sp. Nom.&  {\color{black}  8} & 25.20 & {\color{black} 8} &  {\color{black} 128} &  110& 110 &  {\color{black} 70} &&  {\color{black} 5.04}  & {\color{black} 4.38}   & \\
  27 & L5 & CDFS SWIRE & Sp. Fast & 10 & 41.72 & 20 & 81 & 190 & 150 & 99 &  & 12.18 & 11.39 & \\
  28 & L5 & Lockman SWIRE & Sp. Fast & 2 & 13.51 & 2 & 16 & 218 & 218 & 2 &  & 18.2 & 17.37 & SDP\\
  28B & L5 & Lockman SWIRE & Sp. Fast &  {\color{black} 8} & 41.26 & 8 &{\color{black} 58}  & 220 & 180 & 50 &  & 
  {\color{black}15.26} & {\color{black} 7.63} & \\
  42 & L7 & HeLMS & Sp. Fast & {\color{black} 11(10)} & 103.4 & 2 &  & 1560 &750 &  15 & 270&  &  & \\

  \hline
  20 & L3 & Lockman-North & PACS & 12 & 13.96 & 11 &  & 30 & 30 & 42 & 0.25 &  &  & SDP\\
  20B & L3 & Lockman-North & PACS & 20 & 20.89 & 20 &  & 30 & 30 & 42 & 0.25 &  &  & \\
  21 & L3 & UDS HerMES & PACS & 25 & 25.93 & 14 &  & 30 & 30 & 0 & 0.25 &  &  & \\
   25 & L4 & UDS & PACS & 12 & 40.19 & 7 &  & 57 & 57 & 0 & 0.9 &  &  & \\
 \hline

  29 & L5 & EGS HerMES & Parallel & {\color{black} 7} & 22.68 & 7 & {\color{black}93} & {\color{black} 150} & {\color{black}40} & {\color{black} 131} & & {\color{black}3.50} & {\color{black} 2.67} & \\
  30 & L5 & Bootes HerMES & Parallel & 5 & 20.33 & 5 & 70 & 80 & 80 & 0 & & 4.21 & 3.25 & DR1\\
  31 & L5 & ELAIS N1 HerMES & Parallel & 5 & 20.82 & 5 & 72 & 95 & 95 & 38 & & 3.74 & 3.25 & DR1\\
  32 & L5 & XMM VIDEO1 & Parallel & {\color{black} 4} & 13.44 & 4 & {\color{black} 65} & {\color{black} 106} & {\color{black} 75} & {\color{black} 107} &  & {\color{black} 3.20} & {\color{black} 2.72} & \\
  32B & L5 & XMM VIDEO2 & Parallel & {\color{black} 4} & 8.88 & 4 & {\color{black} 53} & {\color{black} 106} & {\color{black} 44} & {\color{black} 107} &  & {\color{black} 2.12} & {\color{black} 1.74}& \\
  32C & L5 & XMM VIDEO3 & Parallel & {\color{black} 4} & 13.44 & 4 & {\color{black} 53} & {\color{black} 106}  & {\color{black} 75} & {\color{black} 107}  &  & {\color{black} 3.19} &{\color{black}  2.73} & \\
  33 & L5 & CDFS SWIRE & Parallel & {\color{black} 4} & 50.42 & {\color{black} 4} & {\color{black} 57} &  {\color{black} 204} & {\color{black} 170} & {\color{black} 101} &   &  {\color{black} 11.87}  &   {\color{black} 10.89} & \\
  34 & L5 & Lockman SWIRE & Parallel & {\color{black} 4(2)} & 71.22 & 4 &  & {\color{black} 215} & {\color{black} 215} & {\color{black} 154} & & {\color{black}17.86}  &  {\color{black}16.08} & \\
  39B & L5 & ELAIS S1 VIDEO & Parallel & {\color{black} 4} & 17.72 & 4 & {\color{black} 56} & {\color{black} 138} &{\color{black} 80} & {\color{black} 87} &  & {\color{black} 4.42} & {\color{black}3.72} & \\
  35 & L6 & ELAIS N1 SWIRE & Parallel & {\color{black} 2} & 28.0 & 2 & {\color{black} 28} & {\color{black} 207} & {\color{black}192} & {\color{black}55} &  & {\color{black} 13.37} & {\color{black}12.28} & \\
  36 & L6 & XMM-LSS SWIRE & Parallel & 6 & 45.58 & 2 & 29 & 180 & 180 & 82 &  & 21.62 & 18.87 & DR1\\
  37 & L6 & Bootes NDWFS & Parallel & 4 & 27.99 & 2 & 30 & 243 & 80 & 145 &  & 11.3 & 10.57 & DR1\\
  38 & L6 & ADFS & Parallel & 2 & 18.11 & 2 & 28 & 190 & 122 & 80 &  & 8.58 & 7.47 & DR1\\
  39 & L6 & ELAIS S1 SWIRE & Parallel & 2 & 17.9 & 2 & 28 & 140 & 81 & 91 &  & 8.63 & 7.86 & \\
  40 & L6 & FLS & Parallel & 2 & 17.1 & 2 & 29 & 160 & 138 & 5 &  & 7.31 & 6.71 & SDP\\
  41 & L6 & ELAIS N2 SWIRE & Parallel & {\color{black}  2} & 17.1 & 2 & {\color{black} 26} & {\color{black} 177} & {\color{black} 119} & {\color{black}147}  &  & {\color{black} 9.06} & {\color{black}7.80} & \\
\hline\end{tabular}

\caption{Summary of the HerMES observations. The full set of Astronomical Observation Requests (AORs) are available through ESA's {\em Herschel} Archive. We have grouped $N_{\rm AOR}$ observations of the same field at the same level {\color{black} made} with the same mode and areal size into a `set' (the number of AORs still to be {\color{black} scheduled after 2011~Dec~21} is indicated in parentheses). The first five columns in the Table give: the set identification number; the design level; the target name, the {\em Herschel} observing mode and the number of AORs in the set.   $T$ is the time used or allocated for this set.  $N_{\rm rep}$ is the total number of repeats of the observing mode in the set. All our SPIRE nominal {\color{black}(30\arcsec\, s$^{-1}$) and fast mode (60\arcsec\, s$^{-1}$)} (Sp. Nom. and Sp. Fast) observations include a scan in the nominal  and orthogonal direction, so 1 repeat is 2 scans). For SPIRE observations that have been executed $N_{\rm samp}$ is the median number of bolometer samples per pixel in the 250~\micron\ map ($6\arcsec\times 6\arcsec$ pixels).{\color{black} This can be converted to exposure time per pixel or to other bands as described in footnote~2.}
 The error per pixel in our SPIRE maps as processed by the standard HIPE pipeline are $\sigma_{250}^2=\sigma_0^2/N_{\rm samp}$ with $\sigma_0^2=896\pm11, 1554\pm 27$ and $\sim 1440$ ${\rm mJy}^2\, {\rm beam}^{-2}$ for Parallel, Sp. Nom. and Sp. Fast modes respectively. $l_1,l_2$ are sides of a rectangle with near homogenous coverage.  $\theta$ is the roll angle with short-axis of that rectangle measured East of North. For SPIRE observations that have been executed $\Omega_{\rm max}$ is the total area of pixels with any 250$\,$\micron\ coverage and  
$\Omega_{\rm good}$ is the area of pixels where the number of bolometer samples per pixel in the 250$\,$\micron\ map is greater than 
$N_{\rm samp}/2$.  For PACS fields or unobserved SPIRE fields $\Omega_{\rm nom}$ gives the nominal area of region. The final column indicates which observations are included in our data releases; observations marked SDP were released in our Early Data Release, observations marked SDP or DR1 will be released in our First Data Release.  Set numbers \#16 and \#26 were removed from the programme. }\label{tab:AORs}
\end{table*}

\subsection{Requirements}
HerMES was designed to fulfil multiple objectives, which are outlined in Section~\ref{sec:method}.  
 {\color{black} The {\em Herschel} bands can probe the peak of the far infrared spectral energy distributions of star forming galaxies and thus measure the infrared luminosity, $L_{\rm IR}$, see Figure~\ref{fig:sed} and Table~\ref{tab:bands}.}
Our primary criterion was to  sample the $(L_{\rm IR} ,z)$ plane of star-forming galaxies uniformly and with sufficient statistics to a redshift of $0<z\la3$.  Specifically, we take a bin resolution of $\Delta \log L_{\rm IR}\, \Delta z=0.1$ (e.g. $\Delta \log L_{\rm IR}=0.5, \Delta z=0.2$) and require 75 galaxies per bin to give 12 per cent accuracy (or 10 per cent accuracy when further divided into three sub-samples).  This resolution corresponds to the scale of features in the luminosity density surface from the \citet{Lagache:2003} model, for example.  Using the model luminosity functions we can calculate the area needed to reach this goal for each luminosity and redshift.  Each tier thus probes a given $(L_{\rm IR},z)$  region bounded by the areal constraint and the flux limit (see Fig.~\ref{fig:lumden}).  An optimized sampling over wavelength is achieved by combining HerMES with the PACS Evolutionary Probe survey \citep[PEP,][]{pep}.

\begin{table}
{\color{black}
\begin{tabular}{lrrrrr}
  	&    \multicolumn{5}{c}{at Nominal Wavelength  [\micron]} \\
 	&100	&160&	250&	350&	500 \\ \hline
Instrument & PACS & PACS & SPIRE & SPIRE & SPIRE\\
Filter name & Blue2 & Red & PSW & PMW & PLW \\
Min $\lambda$ [\micron] & 85   & 125  & 210 & 300 & 410 \\
Max $\lambda$  [\micron]	& 125 & 210 & 290 & 400 & 610 \\

\end{tabular}
\caption{Basic band information for the different {\em Herschel} channels used by HerMES.  Data is taken from SPIRE and PACS Observers' Manuals V2.4/V2.3 (respectively).}\label{tab:bands}
}
\end{table}

HerMES was thus designed to comprise a number of tiers of different depths and areas (Tables \ref{tab:wcake} and \ref{tab:250counts}).  HerMES samples the higher luminosity objects, which are bright but rare, in the wide shallow tiers, and the lower luminosity galaxies, which are faint but common and confused, in the deep narrow tiers.  Our design has evolved during the mission but since our initial design had cluster observations (nominally deep, shallow and high-$z$) and six nominal levels from deep and narrow Level~1 to wide and shallow Level~6 and we will maintain those descriptions even though the depths have changed. 

Confusion is a serious issue for {\em Herschel} and SPIRE in particular, and is an important driver in deciding survey depth (Table~\ref{tab:wcake}).  
To estimate the confusion level we assembled galaxy models \citep[e.g.][]{Lagache:2003}, compared them to existing survey data,
and calculated the confusion limit using the criteria for source density of 30 beams per source and width of the sky intensity distribution.  We employ a number of techniques to overcome the problem of confusion.  It is those analyses which motivate the deepest tiers: the lensed clusters fields; and the fast scanned elements of the wide Level~5 tier.

An additional consideration is the volume of the survey needed for a representative sample of the Universe, to provide a
sufficient range of environments, and enough independent regions to study clustering (e.g. Fig.~\ref{fig:virgo}).  Examination
of each of those requirements requires survey co-moving volumes of $10^6-10^7 {\rm Mpc}^3$ or larger. E.g. the number density today of  halos of dark matter mass $M_{\rm DM}>10^{15}\,{\rm M}_{\sun}$ is around $10^{-6}\,h^{3}{\rm Mpc}^{-3}$\citep{MoWhite2002}.
This is identical to the co-moving number density of their progenitors i.e. $\sim 0.3-0.4\,{\rm deg}^{-2}$ for survey shells of  $\Delta z=0.1$.  This provides additional motivation for fields of order 10s deg$^2$ to provide statistical samples. Sampling variance would still be an issue if the smaller deeper levels were contiguous so we split these into a number of fields to enable us to reduce and estimate the sampling variance errors.

The SPIRE and PACS depths for the cluster observations were determined by the  
desire to ensure the detection of $z \simeq 3$ sources
in both the {\color{black} SPIRE 250 and PACS} 100$\,$\micron\ channels.  

\subsection{Choice of Fields}\label{Sec:ChoiceFields}

\begin{figure*}
\includegraphics[width=16cm]{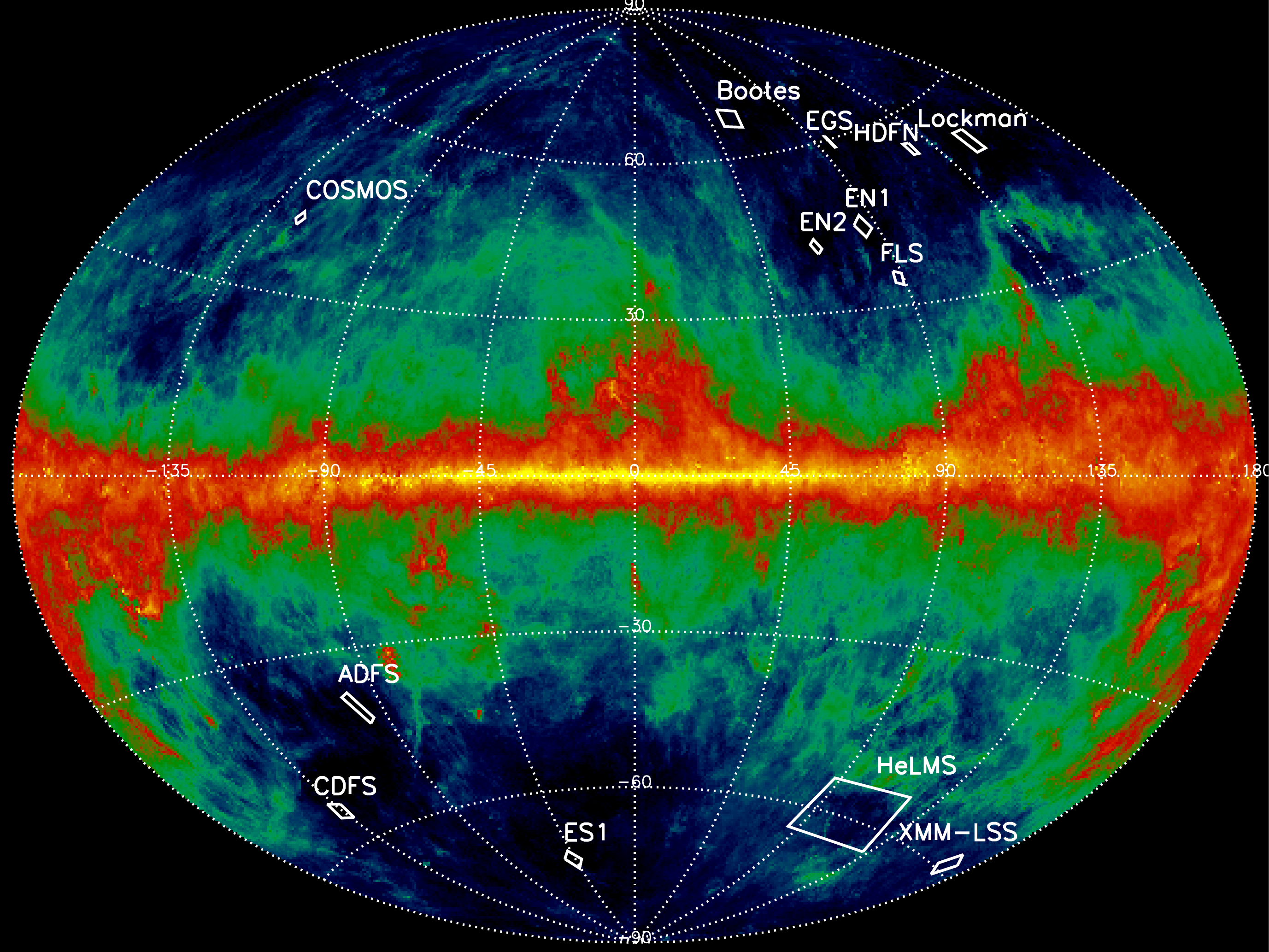}
\caption{Map of dust emission from the Galaxy, with HerMES fields over-plotted. The image is the 100$\,$\micron, {\em COBE}--normalised, {\em IRAS} map of extended emission \protect\citep{Schlegel1998}. The projection is Hammer-Aitoff in Galactic coordinates. The sky brightness is plotted on a false-colour logarithmic scale, with regions of very low Galactic emission appearing black and the Galactic plane yellow. In addition to the blank fields marked, HerMES has also observed 12 known clusters.}\label{fig:fields}
\end{figure*}
  
In order to pursue multi-wavelength analyses,  we have selected fields (Fig.~\ref{fig:fields}, Table~\ref{tab:wcake}) which are among the most intensively observed at all wavelengths.  These incude: radio  (VLA, WRST, GMRT, ATCA); sub-mm (SCUBA, Bolocam, AzTEC, MAMBO); mid and far infrared ({\em Spitzer}, {\em ISO}, {\em AKARI}); near-infrared (UKIRT, VISTA); optical ({\em HST}, Subaru SuprimeCAM, CFHT MegaCAM, KPNO Mosaic1,  CTIO MOSAIC2, INT WFC); UV ({\em GALEX})  and  X-ray ({\em XMM-Newton},{\em Chandra}).  A description of ancillary data is given in Section~\ref{sec:ancillary}).
Extensive redshift and/or photometric redshift surveys are either available or underway for most of these fields.

An additional consideration was that the contamination from Galactic emission (or cirrus) should be minimal.  The larger mirror means that this cirrus is less of a concern for extra-galactic surveys with {\em Herschel} than it was for {\em Spitzer}, as discussed in \citet{Oliver2001}.  This means that our requirement for low-levels of cirrus are automatically satisfied by our criteria of good ancillary data, as illustrated in Fig.~\ref{fig:fields}.

The defining criterion was coverage at mid/far infrared wavelengths not accessible to {\em Herschel}, or where {\em Herschel} is relatively inefficient due to its warm mirror.  Specifically we required  {\em Spitzer} MIPS coverage at 24 and 70$\,$\micron.  At the time of design the one exception to this was the {\em AKARI} Deep Field South, which did not have  {\em Spitzer} coverage but did have coverage at 65, 90, 140, and 160$\,$\micron\ from {\em AKARI} \citep{adfs1}. However, this field has since been observed by  {\em Spitzer} MIPS \citep{adfs2a}. The HeLMS field, which was added in 2011 for studying large-scale structure and the bright end of the number counts, does not have ancillary data from {\em Spitzer}.  However, being located on the SDSS Stripe 82 region, HeLMS does have ancillary coverage from many other facilities.

A detailed discussion of the specific observations which were considered in the design of the fields is given in Appendix~\ref{sec:fieldrational}

The deep and shallow cluster targets are well-studied strong lenses at modest redshift. 
They were selected in consultation with the PEP team -- with HerMES carrying out the SPIRE observations and PEP the corresponding PACS observations.
The high-$z$ clusters were selected for environmental studies also in consultation with the PEP team.

\begin{figure*}
\includegraphics[height=3.5cm]{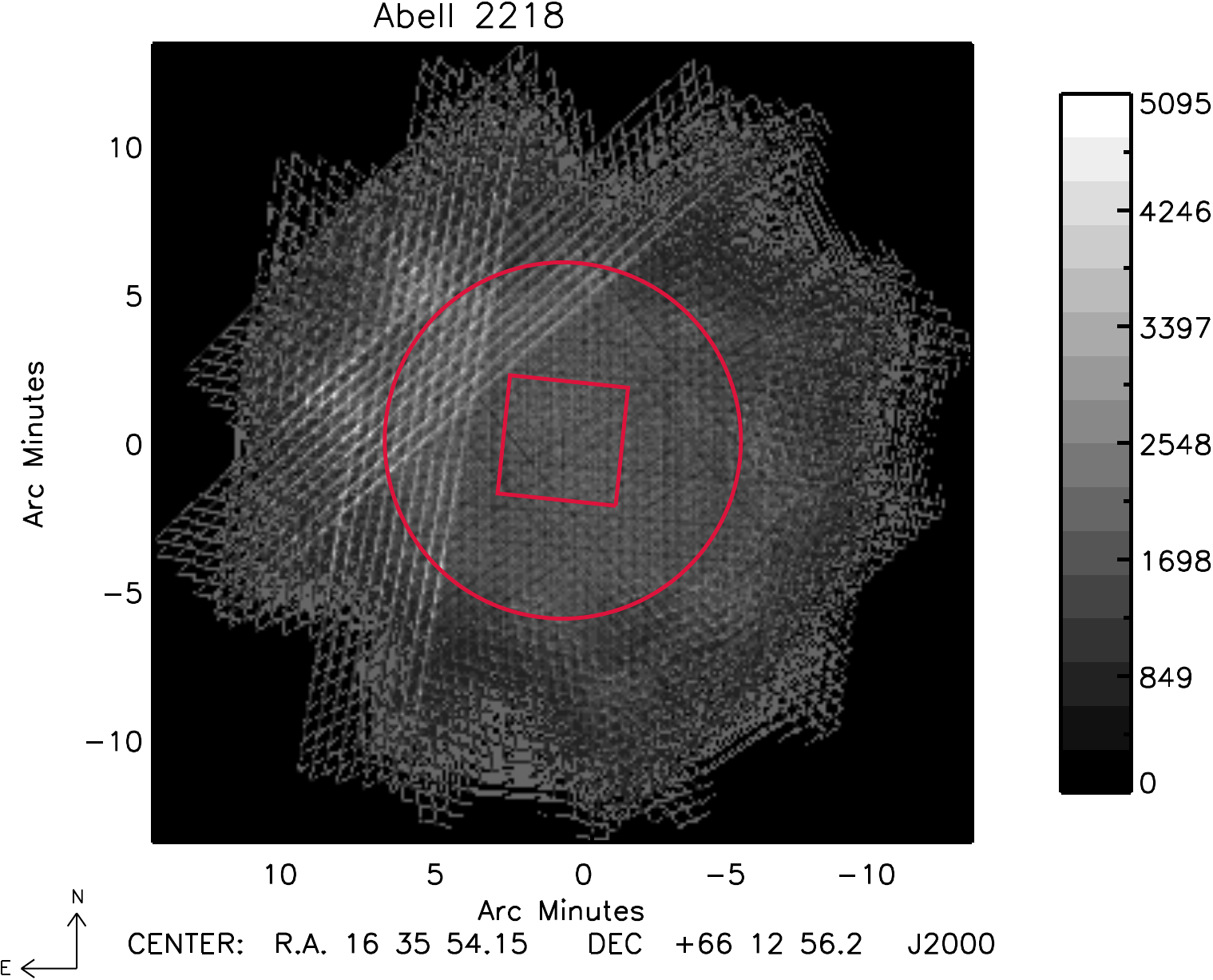}
\includegraphics[height =3.5cm]{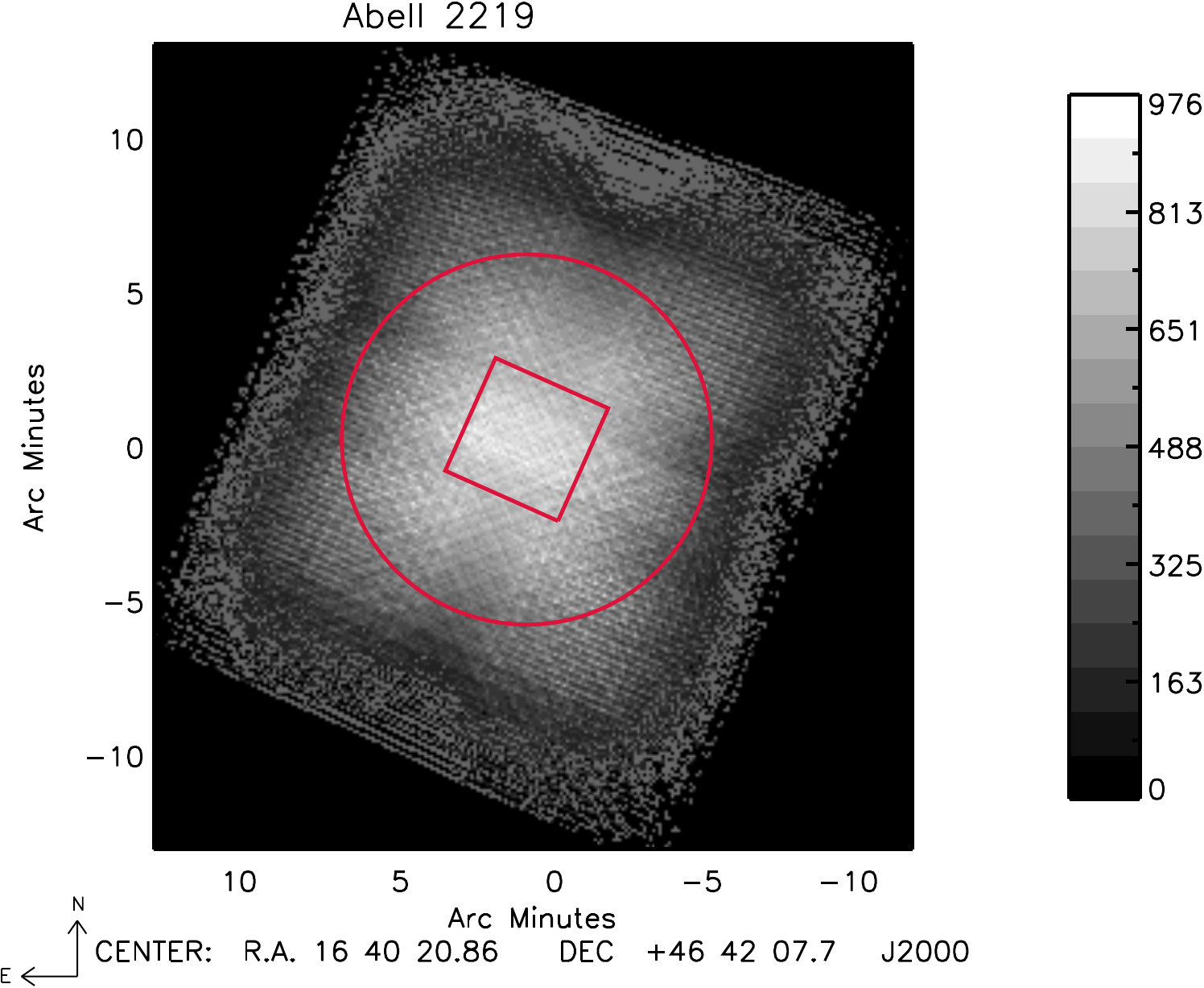}
\includegraphics[height =3.5cm]{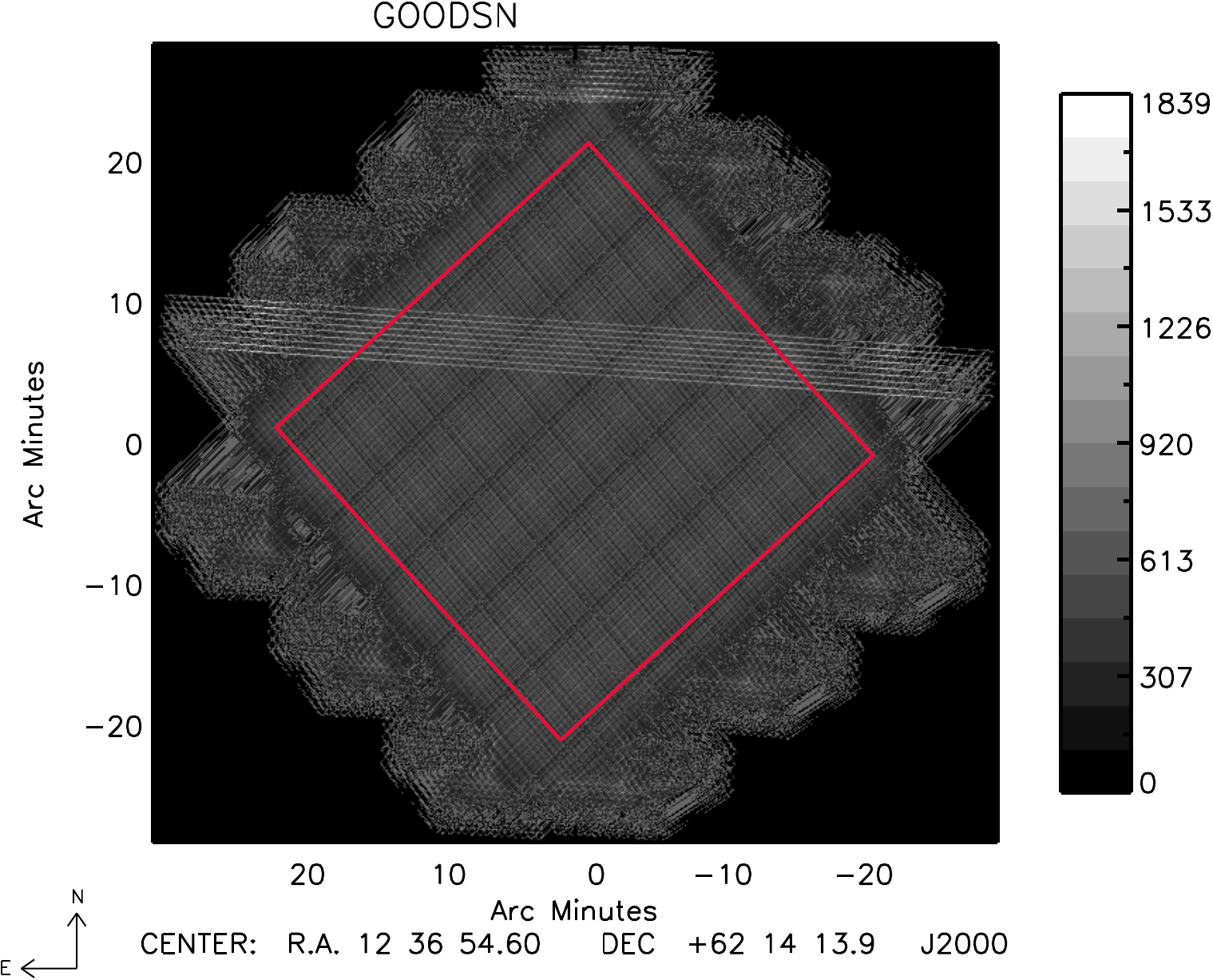}
\includegraphics[height =3.5cm]{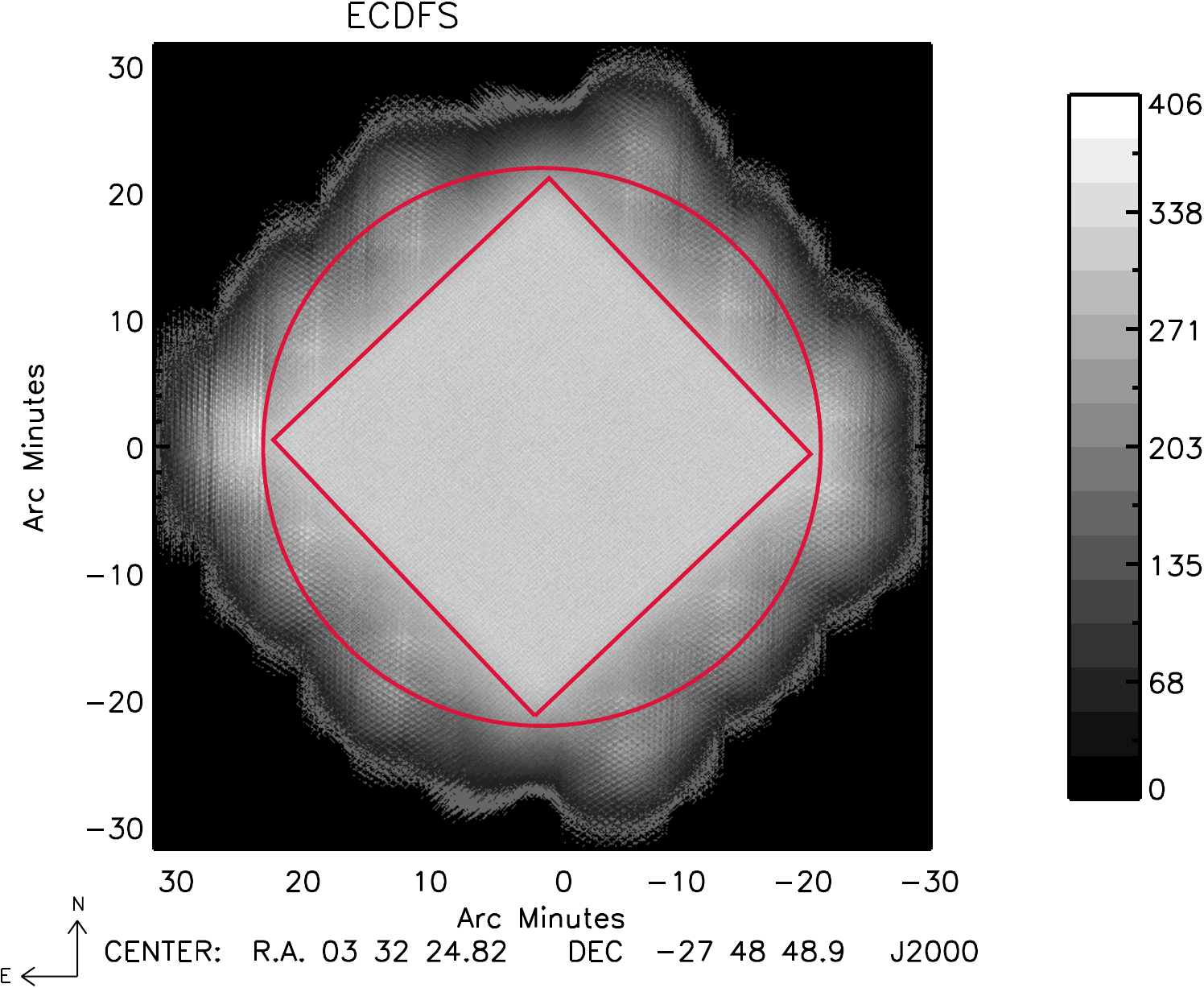}
\caption{Maps of the number of bolometer samples per pixel of four deep SPIRE 250$\,$\micron\ observations. 
From left to right: Abell~2218 which was observed in SDP without dithering; Abell~2219 which was taken with dithering; GOODS-N (taken in SDP without dithering) and ECDFS with dithering.
FITS files of all coverage maps are on http://hermes.sussex.ac.uk/ as will be new coverage maps as the data are taken. }\label{fig:covdither}
\end{figure*}

 \subsection{Observing Modes}\label{Sec:ObsModes}


The mapping of Levels 1-4 (\#1,11-19, 22, 23) is performed using
SPIRE `Large Map' mode. 
{\color{black}This mode is described in detail in the  SPIRE Observers' Manual.}\footnote{{\color{black} The SPIRE ObserversÕ Manual is available from the Herschel Science Centre http://herschel.esac.esa.int/Docs/SPIRE/html/spire\_om.html}}
 This is the default SPIRE observing mode for any field size
larger than 4\arcmin$\times$4\arcmin.  {\color{black} In this mode maps are made by scanning the telescope 
because it eliminates off-beam confusion, allows measurement of extended emission, and increases observing efficiency
for larger fields.  Since our smallest blank field to be mapped (Level~1) is
$\sim$16\arcmin$\times$16\arcmin, {\color{black} this mode} was the natural choice for our program.

The SPIRE cluster observations were originally designed using the `Large Map' mode  covering a nominal field of 4\arcmin$\times$4\arcmin\  {\color{black} as this was the smallest map that could be made using scanning. Abell~2218, \#1, was carried out in that mode}.  We moved to `Small Map' mode  (\#2-10)  {\color{black} in which the map is made by two short cross-scans with the telescope} once that became available, as that was more efficient for small fields.

 {\color{black}  When building maps the telescope} is scanned at an angle {\color{black} of $42.4^\circ$} with respect to the $Z$ axis of the arrays, (see Figure~3.1 and 3.3 of the SPIRE Observers' manual, V2.4). This} produces a fully-sampled map, despite the focal plane not being fully sampled.  The offset between {\color{black} successive scans (or scan `legs') is 348\arcsec,} nearly the full projected array size {\color{black} (see Figure~3.2 of the SPIRE Observers' manual, V2.4)}.  SPIRE observations use two near-orthogonal default scan angles {\color{black}  i.e.  $\pm42.4^\circ$}.  

Multiple map repeats were required to integrate
down to the flux limit in each level. These repeats were
performed with as much cross-linking as possible (i.e. with {\color{black} similar} numbers of scans in quasi-orthogonal directions),
to enable mapping with the presence of {\color{black} low-frequency drifts} and redundancy for the removal of any problematic scans.   
We used the nominal SPIRE scan rate of 30\arcsec${\rm s}^{-1}$ for these fields. 

Where long observations had to be split we aimed to cover the whole field on separate occasions (rather than dividing the field and subsequently building a mosaic) to give redundancy and maximal cross-linking.


The Lockman~SWIRE and CDFS~SWIRE observations in Level~5 (\#27 and 28) were motivated by the study of extragalactic background fluctuations.

These observations required the rapid scanning using the SPIRE fast scan {\color{black} rate} at 60\arcsec${\rm s}^{-1}$  to minimize {\color{black} the effects of low-frequency drifts} and increase redundancy.  The scanning angles and scan leg offsets are the same as for the nominal scan {\color{black} rate}.

{\color{black} The knee frequency is that at which the power of the correlated fluctuations (primarily from the thermal drifts) equates to the white noise.  The design goal for the SPIRE detectors was for the knee to be at 30$\,$mHz (with a requirement of 100$\,$mHz) but the in-flight performance is much better and by using the thermometer signals to de-correlate the drifts knee frequencies of 1-3$\,$mHz can be recovered \citep{Griffin2010}.  The drift is correlated across the detector array (139 bolometers at short wavelengths) and so the effective knee frequency for maps is higher.  Assuming the knee frequency to be 30$\,$mHz thermal drift effects would impact on a spatial scale of 33\arcmin\ (for the fast scan rate) compared to 17\arcmin\ for the nominal scan rate. }



Levels 5 and 6 (\#29--41 and 22B) are being mapped with the SPIRE-PACS parallel mode. {\color{black} This mode is described in detail in the  SPIRE-PACS Parallel Mode Observers' Manual.}\footnote{\color{black} The SPIRE-PACS Parallel Mode Observers' Manual is available from the Herschel Science Centre http://herschel.esac.esa.int/Docs/PMODE/html/parallel\_om.html}
Parallel mode maps the sky simultaneously with both instruments.  {\color{black} The SPIRE detector sampling rate is reduced from 18.2$\,$ Hz to 10$\,$ Hz in this mode, which has a negligible impact when scanning in the slow (20\arcsec${\rm s}^{-1}$) mode}. 
The PACS instrument In the blue channel we used the PACS Blue2 85--{\color{black}125}$\,$\micron\ filter (rather than the 60--85) for maximum sensitivity. We used the 20\arcsec${\rm s}^{-1}$ scanning mode as the 60\arcsec${\rm s}^{-1}$ mode was not suitable for PACS {\color{black} as the beam is degraded by up to 30 per cent \citep[and Table~\ref{tab:beam}]{Poglitsch2010}}. 

 {\color{black} The parallel mode} achieves the
combined PACS and SPIRE sensitivities more efficiently for large areas than observations using each instrument in turn.
Scan directions alternate between nominal and orthogonal for maximal cross linking. 
 
The Level~7, HeLMS, observations (\#42) exploited the ability of the SPIRE to make long (20 deg) scans at the {\color{black} fast (60\arcsec${\rm s}^{-1}$) scan rate}.  These were interleaved in a cross-like configuration to give duplicate coverage in a near-orthogonal direction.  The resulting $270~{\rm deg}^2$ maps are thus optimised for studying fluctuations on the largest possible scale. 
 

All PACS-only observations  (Levels 3--4, \#20, 21, 25, 26) were carried out using {\color{black} the scan mapping mode. This mode is described in detail in the  PACS Observers' Manual.}\footnote{\color{black} The PACS ObserversÕ Manual is available from the Herschel Science Centre http://herschel.esac.esa.int/Docs/PACS/html/pacs\ om.html}

{\color{black} The noise of the PACS bolometer/readout system has a
strong $1/f$ component \citep{Poglitsch2010} and observations need to be modulated on a time-scale of 1-5$\,$Hz.  
We used the 20\arcsec${\rm s}^{-1}$ scan rate in which the  beam has {\sc FWHM}  $\sim$6.8\arcsec\ or $\sim$11.3\arcsec\ in the two bands we use (see Table~\ref{tab:beam}), i.e. sources are modulated on $\sim$2-3$\,$Hz time-scale.  Faster scan rates (e.g. 60\arcsec${\rm s}^{-1}$ in parallel mode) would have introduced significant beam smearing of around 30 per cent \citep[and Table~\ref{tab:beam}]{Poglitsch2010}.}

We alternated orthogonal scan directions to minimise correlated noise, i.e. correlations arising from asymmetric transient detector responses to sky signal. 

\subsection{Dithering}
Moving the array on successive scans  so that different pixels or bolometers trace different parts of the sky (dithering) improves the quality of the data in a number of ways.  Dither steps of more than one detector will reduce correlated noise arising when the same detector crosses the same patch of sky on a short timescale.  Dithering on large scales will also increase uniformity by distributing dead/noisy pixels across the maps. Dithering at sub-detector scales can possibly lead to some improvement in resolution if the point spread function is not fully sampled and (in the case of SPIRE) further reducing the impact of the sparse filling of the focal plane.

For PACS-only observations we implemented a dithering pattern. {\color{black} For each scan we requested an offset with respect to our nominal target position with offsets defined} on a grid with spacing $(0\arcsec, \pm7.5\arcsec, \pm10.5\arcsec)$. This provides sampling at sub-pixel and sub-array scales.

For SPIRE we {\color{black} modelled the}  scan pattern of good detectors and investigated {\color{black} dithering patterns that reduced the variation in sensitivity to point sources (for details see Appendix~\ref{appendix:dither}).} We found that for a given number of repeats, $N$, offsetting by a fraction ${1}/{N}$ of the scan leg separation between repeats was usually close to optimal.  Exceptions to this would be cases where the resulting step size coincided with the projected bolometer spacing, however, none of our patterns resulted in that coincidence.   This also provided a good de-correlation of the noise. The disadvantage of these large dither steps is that the coverage declines at the edges of the map. However, for our large maps this is not a major penalty.  Since each SPIRE-only observation consisted of two scans one at each of the near-orthogonal SPIRE scan angles we set an offset in both directions at once.  We arranged these offset pairs in a square pattern to minimise the edge effects. This dithering was not done for observations taken during the Science Demonstration Phase, but was implemented afterwards.
The contrast in the coverage maps between dithering and not dithering can be seen in Fig.~\ref{fig:covdither}.

\subsection{Sensitivity}\label{Sec:ObsTime}
To estimate the sensitivity of our survey design we use the {\em Herschel} Observation Planning Tool, HSPOT v5.1.1.  For our survey scanning patterns we compute the 5$\sigma$ instrument sensitivity (ignoring confusion noise).  The HSPOT sensitivities are tabulated in Table~\ref{tab:sense} and their implications for Herschel surveys in Table~\ref{tab:wcake}.

\subsection{Economies from nesting}
We have designed our survey starting at the widest, shallower tier and building up the deeper tiers.   Thus a small field tier nested within a shallower tier needs fewer repeats to reach the required depth.  This improves the overall survey efficiency, because observations of small fields are relatively inefficient due to the overheads associated with telescope turn-arounds.



 
 The current coverage of the nested fields around CDFS is illustrated in Fig.~\ref{fig:covcdfs}.

The nesting of fields is {\color{black} indicated in columns 5 and 6 in Table~\ref{tab:wcake}.   
and} the sensitvities in Table~\ref{tab:wcake} take this into account. E.g. UDS-HerMES at Level~3 (\#21) includes 12 PACS scans from UDS Level~4 (\#25), in addition to the 25 from Level~3, giving a total of 37 as well as 14 SPIRE nominal scans from UDS Level~4 (\#23), four Parallel scans from XMM-VIDEO at Level~5 (\#32) and two Parallel scans from Level~6 XMM-LSS~SWIRE (\#36).

\subsection{Total Time}

The total time allocated for HerMES is 909.3 hours. This comes from the Guaranteed Time awarded to the SPIRE instrument team (850 hr) one of the {\em Herschel} Mission Scientists (M. Harwit, 10 hr) and members of the {\em Herschel} Science Centre (B. Altieri, L. Conversi, M. Sanchez Portal and I. Valtchanov, 40hr). ESA also effectively contributed 9.3 hours as we agreed for our Abell 2218 observations in Science Demonstration Phase to be made public immediately and so were not charged for these observations.

\subsection{\color{black}Special requirements and constraints}\label{sec:constraints}
The {\em Herschel} observatory is performing very close to specifications and our survey design is very similar to the one proposed. However, some changes and compromises have been made on the basis of post-launch experience.   

\begin{figure*}
\includegraphics[height=4.5cm]{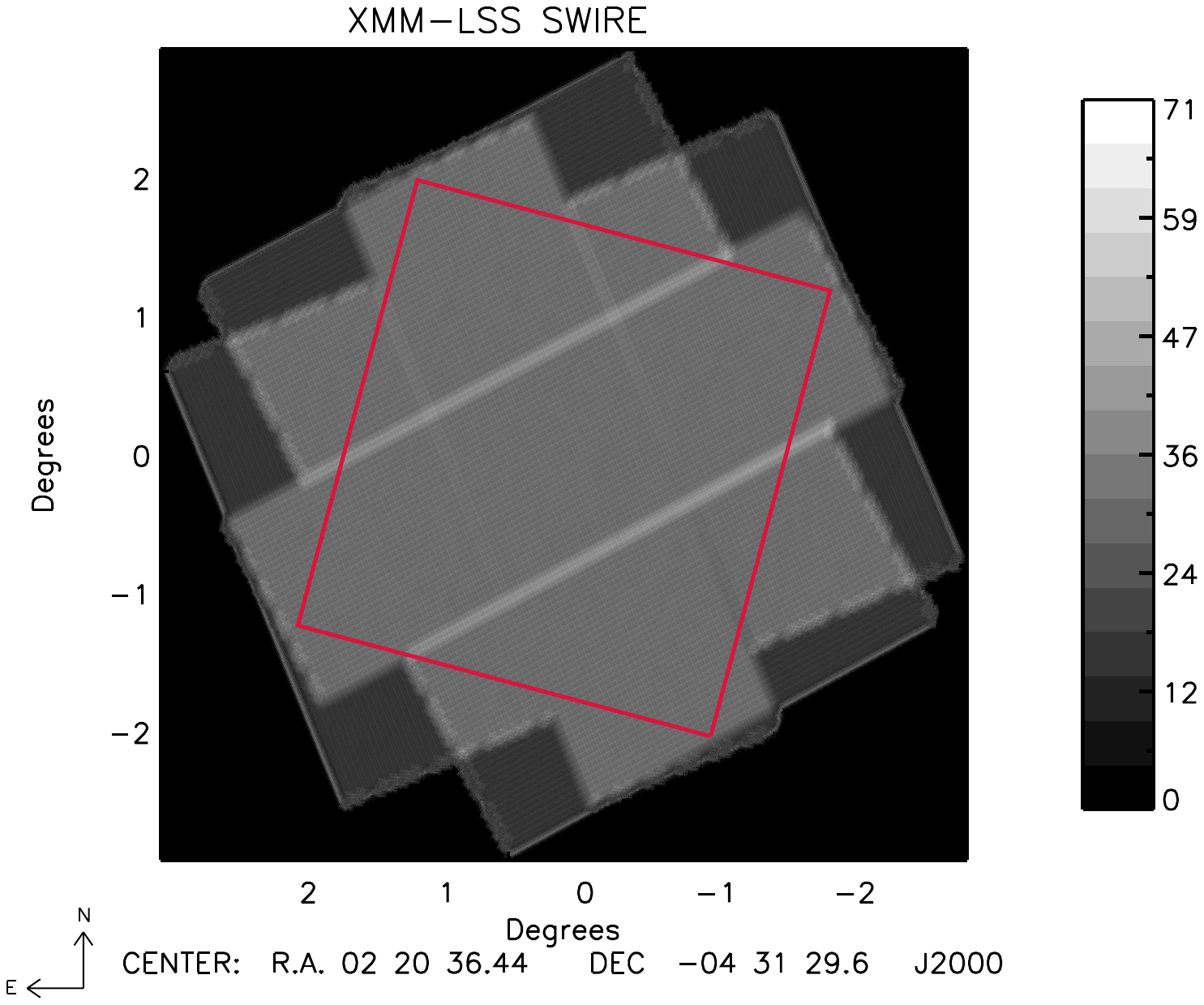}
\includegraphics[height=4.5cm]{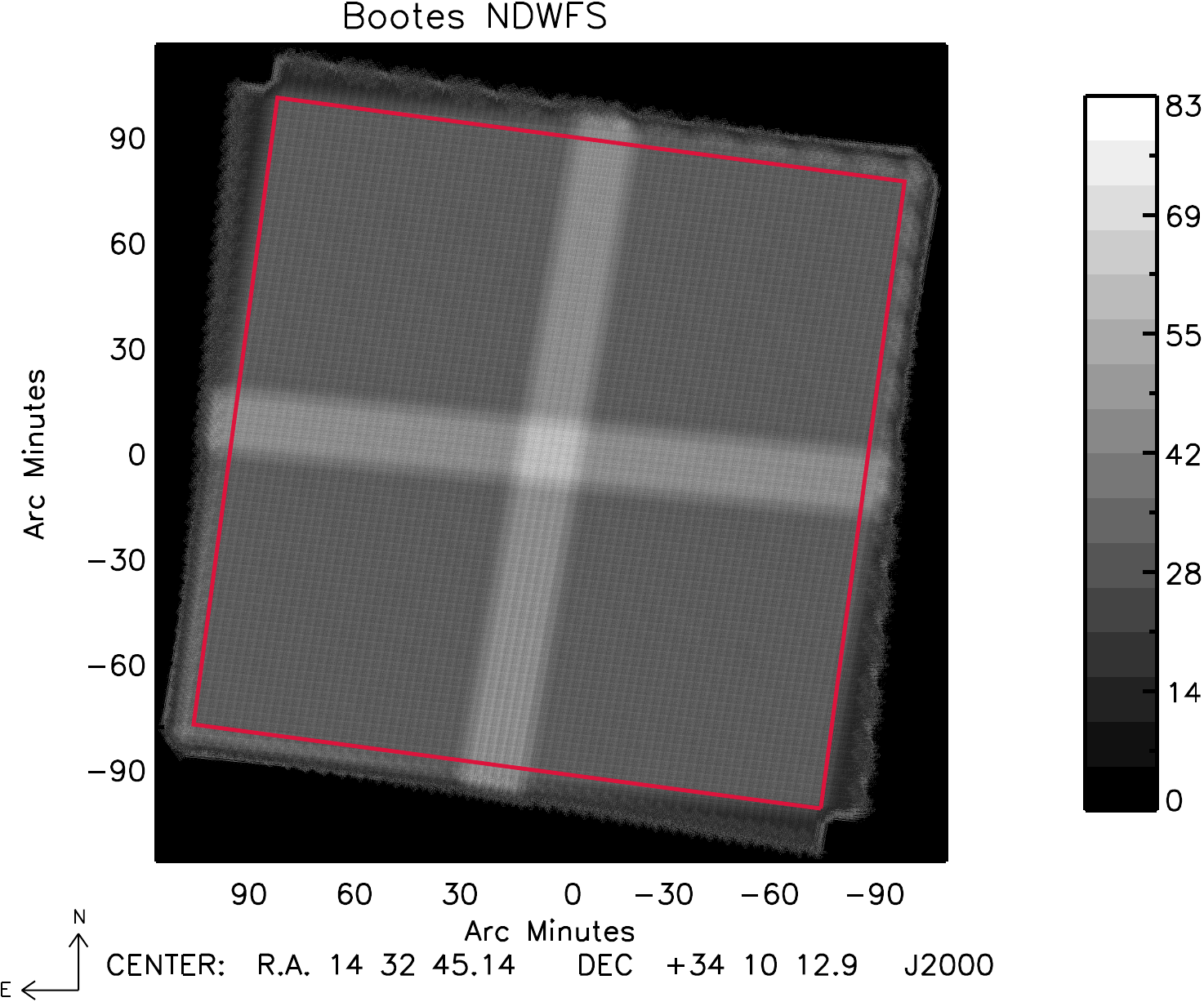}
\includegraphics[height=4.5cm]{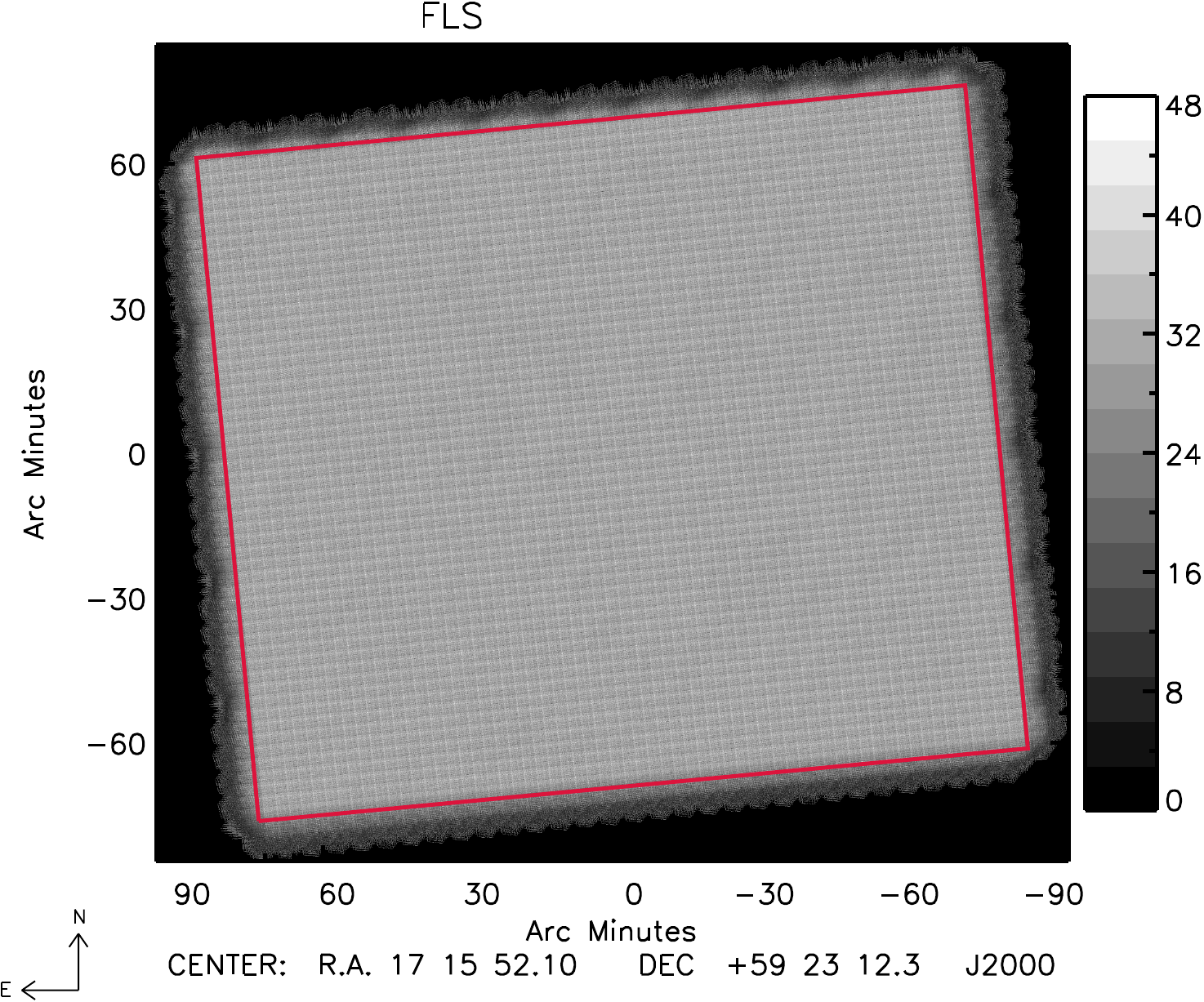}
\caption{Maps of  the number of bolometer samples per pixel in SPIRE 250$\,$\micron\ blank field observations from Level~6. From the left they are   XMM-LSS-SWIRE (\#39), Bo\"otes NWDFS (\#37) taken early with conservative overlap) and FLS (\#40, from SDP).  All are parallel mode  observations with a nominal coverage of two scans. Overlaps produce a maximum coverage of four scans in XMM-LSS-SWIRE and eight in Bo\"otes.
}\label{fig:covmono}
\end{figure*}

\begin{figure}
\includegraphics[width=8cm]{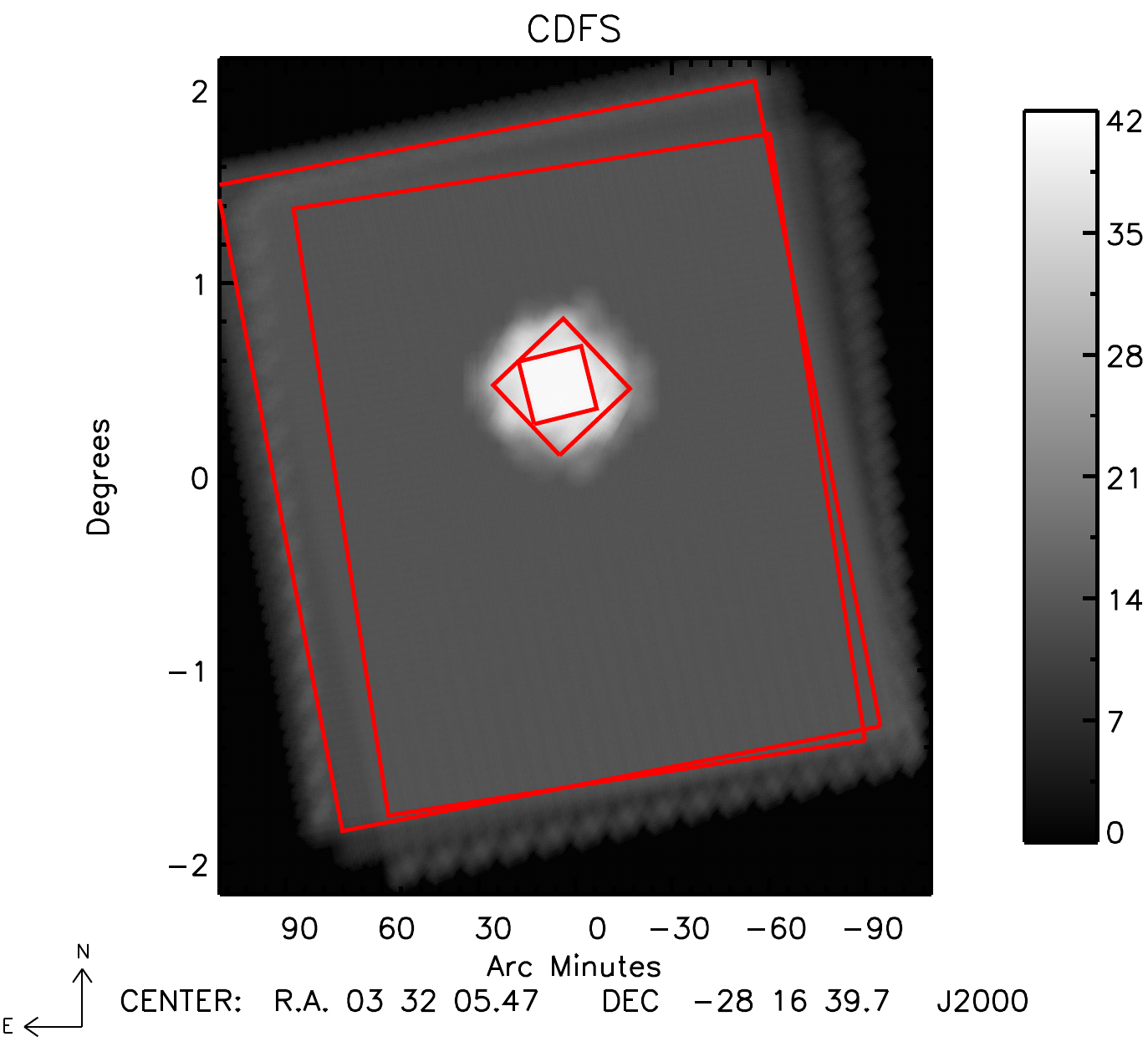}
\caption{Map of square root of number of {\color{black} effective number of} bolometers samples per pixel for  SPIRE 250$\,$\micron\ blank field observations of the CDFS region, which includes our GOODS-S, ECDFS and CDFS-SWIRE observations {\color{black} (\#13,15,27,33). The parallel mode samples (\#33) have been scaled by the relative sampling rates, 18.6/10, to give the effective number of samples they would have had if the observation had been carried out with SPIRE large-map mode with the same exposure time per pixel}.  A region of uniform coverage for each of the {\color{black} independent sets} is marked with a rectangle. {\color{black} N.B. the total coverage drops off in the north-eastern corner of the largest rectangle (delimiting \#33) due to the coverage coming from the boundaries of the large-map mode observations (\#27) but is uniform in a coverage map built from \#33 data alone.}}\label{fig:covcdfs}
\end{figure}

Early in the mission there was a constraint that parallel mode observations could not exceed $2^{15}\,$s, as this exceeded the limit of one software counter.  Since each parallel mode observations was already a single-scan they were as shallow as could be done at that scan rate so this required us to split some of the Level~5 and 6 fields into smaller fields, compromising the uniformity of the data.  
The impact of this on the coverage for the XMM-LSS and Bo\"otes fields is shown in Fig.~\ref{fig:covmono}.  
The planned {\em AKARI} deep field south (\#41) and ELAIS~S1 (\#39) fields required only slightly more time than $2^{15}$~s, and so we chose to reduce the field size rather than split the field.

Where the orientation of the SPIRE data with respect to complementary data was particularly important we constrained the observations to align with them.  Solar avoidance constraints meant that it was not possible to align the SPIRE observations of XMM-LSS~SWIRE (\# 36) and COSMOS optimally with the {\em Spitzer} data and PEP data, respectively. For XMM-LSS~SWIRE we observed a larger field containing the {\em Spitzer} data, while for COSMOS we observed a larger shallower field, COSMOS~HerMES (\#22B), containing the planned PEP PACS observations and a smaller deeper field (COSMOS, \#22), which does not fully cover the PACS observations.  

The  {\em Spitzer}-SERVS and VISTA-VIDEO surveys were approved after HerMES and designed with reference to HerMES.  So, almost all the SERVS and VIDEO fields were included in our Level~5 observations.  However, the SERVS and VIDEO field in ELAIS S1 was not quite within our planned observations, which were only at Level~6.  We thus included additional deeper observations covering the SERVS/VIDEO field (\#39B). 

Our initial Science Demonstration Phase (SDP) observations of Abell~2218 used `Large Map' mode as this was the only way of doing scan mapping. We changed our deep cluster observations to the `Small Map' mode once the mode was available.  

The $P(D)$ results of \citet{Glenn2010} successfully probed the number counts well below the confusion limit, reducing the motivation for exceptionally deep cluster observations.  We have thus reduced the number of repeats.

Due to an error in entering the AOR one parallel observation scan of ELAIS~S1~SWIRE (\#39) was accidentally observed with the shorter wavelength 60--85$\,$\micron\ channel rather than the 85--125$\,$\micron\ channel.

The PACS sensitivity  {\color{black} of 10$\,$mJy (5-$\sigma$ in 1 hr) in the 85--125 channel was significantly less than the   pre-launch estimate (5.3$\,$mJy, PACS Observers' manual v1.1)}  and we removed our planned PACS observations of the VVDS field (\# 26).  

To extend the fluctuation science goals and increase the
{\em Herschel} discovery space for rare objects including gravitationally lensed systems, we added the 
HeRMES Large-Mode Survey (HeLMS), a wide, SPIRE only, tier of $270~{\rm deg}^2$ taking around 100 hours. This exploits
the ability of SPIRE to cover wide areas close to the confusion limit. This additional level is indicated in Table~\ref{tab:wcake}.

\begin{table}
{\color{black}
\begin{tabular}{lllrrrrr}
 &   &  		& \multicolumn{5}{c}{5-$\sigma$ sensitivities [mJy $\sqrt{\rm N_{\rm scan}}$]} \\

Mode &  Rate & Step 		& \multicolumn{5}{c}{at Wavelength  [\micron]} \\
   & [$\arcsec{\rm s}^{-1}$] &  	[\arcsec]	&100	&160&	250&	350&	500 \\ \hline

SPIRE & 	30	& 348 &&	&	64&	53&	76\\ 
SPIRE &	60 	& 348	&&	&	91&	75&	108\\
PACS  & 20 &  55 &		42	&	80	&&	\\	
Parallel  & 20&	168/155	&71	&135&	37&	30&	44 \\  \hline
Parallel  & 60&	168/155 &122&232&	63&	53&	75\\ \hline

\end{tabular}
\caption{Point source sensitivities for different {\em Herschel} observing modes.  Scan rates are given for each mode, we also tabulate the step size between successive scan legs (pre-determined for SPIRE and parallel mode but user-defined for PACS). In parallel mode the step size are different for maps built by scanning in each of the two ``orthogonal'' directions.  5-$\sigma$ sensitivities in units of [mJy $\sqrt{\rm N_{\rm scan}}$]
 for a single scan  are estimated from the HSPOT v5.1.1. Modes below the line are not used by HerMES but by other Key Program surveys.}\label{tab:sense}
}
\end{table}

\begin{table}
{\color{black}
\begin{tabular}{llrrrrr}
  &  		& \multicolumn{5}{c}{Beam {\sc FWHM}  [\arcsec]} \\
Mode  	& Rate	& \multicolumn{5}{c}{at Wavelength  [\micron]} \\
   & [$\arcsec{\rm s}^{-1}$] 	&100	&160&	250&	350&	500 \\ \hline

SPIRE & 	30/60	 &     &&18.2 & 24.9 &  36.3 \\                      
PACS  & 20  &	           6.8  & 11.4 \\    
Parallel  & 20	&   6.8  & 11.4 &     18.2 & 24.9 &  36.3 \\  \hline
Parallel  & 60	& 7.0$\times$12.7 & 11.6$\times$15.7 &              18.2 & 24.9 &  36.3 \\ \hline

\end{tabular}
\caption{Beam sizes for different {\em Herschel} observing modes.  Scan rates are given for each mode. The {\sc FWHM}  of the beams in units of [\arcsec] are taken from SPIRE and PACS Observers' Manuals V2.4/V2.3 (respectively). Where two values are given these are the major and minor axes, when the ellipticity is less than 15 per cent the geometric mean of the two is quoted. The SPIRE beam is not known to vary significantly with scan rate and only one value is given. Modes below the line are not used by HerMES but by other Key Program surveys.}\label{tab:beam}
}
\end{table}

\begin{table*}
{\color{black}
\begin{tabular}{|l|r|r|r|l|l|r|r|r|r|r|r|}
\hline
  \multicolumn{1}{l}{} &
  \multicolumn{3}{l}{Area} &
  \multicolumn{2}{l}{Observations} &
  \multicolumn{5}{l}{5-$\sigma$ noise level (for band in \micron)}  \\
  \multicolumn{1}{|l|}{Fields} &
  \multicolumn{1}{l|}{Nominal} &
  \multicolumn{1}{l|}{Extra} &
  \multicolumn{1}{l|}{Cummulative} &
  \multicolumn{1}{l|}{PACS} &
  \multicolumn{1}{l|}{SPIRE} &
  \multicolumn{1}{c|}{110} &
  \multicolumn{1}{c|}{160} &
  \multicolumn{1}{c|}{250} &
  \multicolumn{1}{c|}{350} &
  \multicolumn{1}{c|}{500} \\
\multicolumn{1}{|l|}{} &
  \multicolumn{3}{c}{[deg$^2$]} &
  \multicolumn{2}{c}{} &
  \multicolumn{5}{c}{[mJy]}  \\

\hline
  Abell 2218 & 0.0050 & 0.0050 & 0.1 & P & 1 & 4.1 & 7.9 & 6.4 & 5.3 & 7.6\\
  Abell 1689 & 0.0050 & 0.0050 & 0.11 & P & 2 & 3.6 & 6.9 & 9.2 & 7.7 & 11.0\\
  8 Targets & 0.04 & 0.04 & 0.15 & P & 3-10 & 5.7 & 10.9 & 9.2 & 7.7 & 11.0\\
  2 Targets & 0.03 & 0.03 & 0.18 & P & 11-12 &  &  & 13.9 & 11.6 & 16.7\\
  Various & 0.18 & 0.18 & 0.36 & E & E & 6.1 & 11.7 & 14.2 & 11.9 & 17.1\\ \hline
  GOODS-N & 0.042 & 0.042 & 0.04 & G,P & G,14 & 2.2 & 4.1 & 3.8 & 3.1 & 4.5\\
  GOODS-S & 0.11 & 0.087 & 0.13 & G,P,33 & 13,15,27,33  & 2.1 & 2.9 & 4.3 & 3.6 & 5.2\\
  GOODS-S & 0.012 & 0.012 & 0.14 & G,P,33 & 13,15,27,33  & 1.1 & 2.1 & 4.6 & 3.8 & 5.5\\
  GOODS-S & 0.018 & 0.0060 & 0.15 & G,P,33 & 13,15,27,33  & 1.6 & 3.0 & 4.6 & 3.8 & 5.5\\
  GOODS-S & 0.023 & 0.0060 & 0.15 & G,P,33 & 13,15,27,33  & 2.0 & 3.8 & 4.6 & 3.8 & 5.5\\
  COSMOS & 2.0 & 2.0 & 2.15 & P & 22,22B & 7.7 & 14.7 & 8.0 & 6.6 & 9.5\\
  ECDFS & 0.25 & 0.14 & 2.29 & P,33 & 15,27,33  & 7.6 & 14.5 & 8.0 & 6.6 & 9.6\\
  GOODS-N & 0.25 & 0.208 & 2.5 & P & 14 & 4.7 & 8.9 & 8.2 & 6.8 & 9.9\\
  Lockman-East & 0.25 & 0.25 & 2.75 & P & 18,18B,28B,34,28 & 6.5 & 12.3 & 9.6 & 7.9 & 11.5\\
  Lockman-North & 0.25 & 0.25 & 3.0 & 20,20B,34 & 19,28B,34,28 & 7.4 & 14.1 & 10.6 & 8.8 & 12.7\\
  Groth Strip & 0.25 & 0.25 & 3.25 & P,29 & 17,29 & 7.1 & 13.6 & 10.7 & 8.9 & 12.8\\
  UDS HerMES & 0.25 & 0.25 & 3.5 & 21,25,32,36 & 23,25,32,36 & 6.8 & 12.9 & 11.2 & 9.3 & 13.4\\
  UDS & 0.7 & 0.7 & 4.2 & 25,32,36 & 25,32,36 & 11.2 & 21.4 & 11.2 & 9.3 & 13.4\\
  VVDS & 2.0 & 2.0 & 6.2 & 25,32C,36 & 25,32C,36 & 28.8 & 54.9 & 11.2 & 9.3 & 13.4\\ \hline
  CDFS SWIRE & 11.4 & 11.1 & 17.3 & 33 & 27,33  & 31.5 & 60.2 & 12.7 & 10.5 & 15.2\\
  Lockman SWIRE & 16.1 & 15.6 & 32.9 & 34 & 28,28B & 35.3 & 67.3 & 13.6 & 11.2 & 16.2\\
  EGS HerMES & 2.7 & 2.5 & 35.4 & 29 & 29 & 26.6 & 50.8 & 13.8 & 11.3 & 16.4\\
  Bo\"otes HerMES & 3.3 & 3.3 & 38.6 & 30,37 & 30,37 & 26.6 & 50.8 & 13.8 & 11.3 & 16.4\\
  ELAIS N1 HerMES & 3.3 & 3.3 & 41.9 & 31,35 & 31,35 & 26.6 & 50.8 & 13.8 & 11.3 & 16.4\\
  ELAIS S1 VIDEO & 3.7 & 3.7 & 45.6 & 39B,39 & 39B,39 & 28.8 & 54.9 & 14.9 & 12.2 & 17.8\\
  XMM-LSS VIDEO & 7.7 & 5.0 & 50.6 & 32,32B,32C,36 & 32,32B,32C,36 & 28.8 & 54.9 & 14.9 & 12.2 & 17.8\\
  COSMOS Hermes & 4.4 & 2.4 & 53.0 &  & 22B &  &  & 15.9 & 13.3 & 19.1\\
  ELAIS N2 SWIRE & 7.9 & 7.9 & 60.9 & 41 & 41 & 49.9 & 95.1 & 25.8 & 21.2 & 30.8\\
  FLS & 6.7 & 6.7 & 67.6 & 40 & 40 & 49.9 & 95.1 & 25.8 & 21.2 & 30.8\\
  ADFS & 7.5 & 7.5 & 75.1 & 38 & 38 & 49.9 & 95.1 & 25.8 & 21.2 & 30.8\\
  ELAIS S1 SWIRE & 7.9 & 4.2 & 79.2 & 39 & 39 & 49.9 & 95.1 & 25.8 & 21.2 & 30.8\\
  ELAIS N1 SWIRE & 12.3 & 9.1 & 88.3 & 35 & 35 & 49.9 & 95.1 & 25.8 & 21.2 & 30.8\\
  Bo\"otes NDWFS & 10.6 & 7.3 & 95.6 & 37 & 37 & 49.9 & 95.1 & 25.8 & 21.2 & 30.8\\
  XMM-LSS SWIRE & 18.9 & 15.0 & 110.6 & 36 & 36 & 49.9 & 95.1 & 25.8 & 21.2 & 30.8\\
  Various & 570.0 & 570.0 & 681.0 & A & A & 86.3 & 164.0 & 44.5 & 37.1 & 53.0\\
  SPT & 100.0 & 100.0 & 781.0 &  & S &  &  & 45.3 & 37.5 & 54.1\\
  HeLMS & 270.0 & 270.0 & 1051.0 &  & 42 &  &  & 64.0 & 53.0 & 76.5\\
  \hline\end{tabular}

}
\caption{{\color{black} HerMES survey with sensitivities in the context of other survey programmes being undertaken by {\em Herschel}.  The ``observations'' columns refer to the AOR set numbers of  Table~\protect\ref{tab:AORs} for HerMES or for other Key Programmes we use: ``E'' for Egami cluster programme, ``G'' for GOODS-H, ``P'' for PEP, ``A" for H-ATLAS and ``S" for SPT (see Table \protect\ref{tab:herschelprogs}).  The sensitivities are estimated consistently using HSPOT v5.1.1.  These are single pixel sensitivities and ignore the benefits of matched filters, particularly for unconfused fields, e.g.  H-ATLAS quote empirical 5-$\sigma$ sensitivities of 105, 140, 32, 36, 45~mJy for the five wavelengths so the sensitivities in this Table should be scaled by 1.22, 0.85, 0.72, 0.97, 0.85 to obtain a consistent comparison with H-ATLAS. The sensitivity of HerMES observations have been calculated including data from shallower tiers as described in the text. Other surveys are treated independently.  Cluster observations are listed before blank fields. The fields are ordered in increasing 250$\,$\micron\ flux limit then right ascension.   The area is defined by the PACS observations for Levels~1-4 (above the second horizontal line), otherwise we use $\Omega_{\rm good}$ from Table~\protect\ref{tab:AORs} or $\Omega_{\rm nom}$ for HeLMS. We tabulate three areas: the nominal area for each field; the `doughnut' area which excludes any deeper sub-fields within; and the cumulative area of all fields higher in the table. The 5-$\sigma$ confusion noise (after $5\sigma$ cut) from \protect\cite{Nguyen2010} is  $24.0, 27.5, 30.5$~mJy (at 250, 350 and 500$\,$\micron), approximately the Level~6 depth. GOODS-S also has PACS data not listed here at 70\micron\ over 0.11$\,$deg$^2$ to a 5-$\sigma$ depth of 1.9$\,$mJy.}}\label{tab:wcake}
\end{table*}


\subsection{Observations}
\begin{figure*}
\includegraphics[width=16cm]{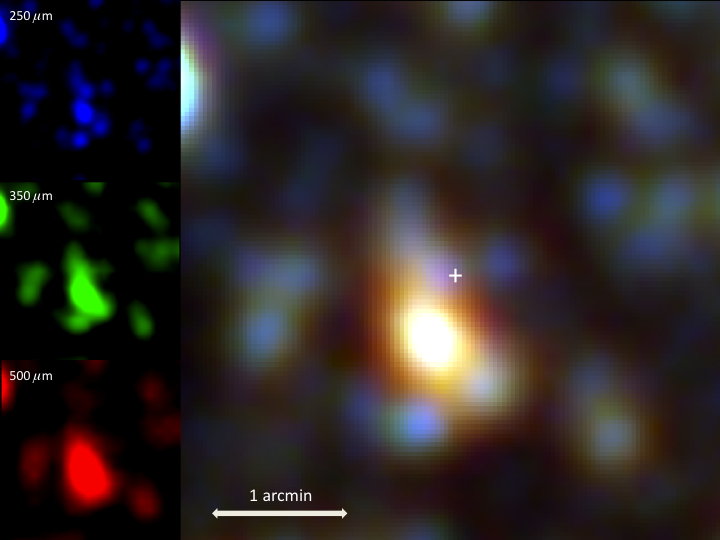}
\caption{Three colour {\em Herschel}-SPIRE image of the central $4\arcmin\times4\arcmin$ of the galaxy cluster Abell 2218.  The left-most panels show the single band images of the cluster, while the central panel 
shows a three colour image generated by resampling the single band images and their flux scalings to a common pixelization.  The centre of the cluster is marked with the cross hairs and a 1\arcmin\ bar is shown for scaling; north is toward the top of the page.  The orange object to the south-east and white object to the south-west of the cluster are images of the multiply imaged sub-mm source studied in detail by e.g.~\protect\citet{Kneib:2004}; this source has been identified to lie at $z=2.516$ though due to the complex mass structure of this cluster each image is magnified by a different amount.  In the SPIRE bands this source's integrated flux densities are measured to be $\{170, 197, 231 \}\,$mJy, corresponding to background flux densities of $\{11.7, 13.5, 15.4 \} \,$mJy. The varying colour of the images suggests that different regions of the source galaxy are being imaged to different points in the map.  In addition, the known $z=4.04$ sub-mm source is seen as the pink object just to the east of the cross hairs \protect\citep{Knudsen:2009}.  The other objects scattered through the image are more typical $z \sim 1$ sources with SEDs peaking shortward of 250$\,$\micron.}\label{fig:a2218}
\end{figure*}

\begin{figure*}
\includegraphics[width=16cm]{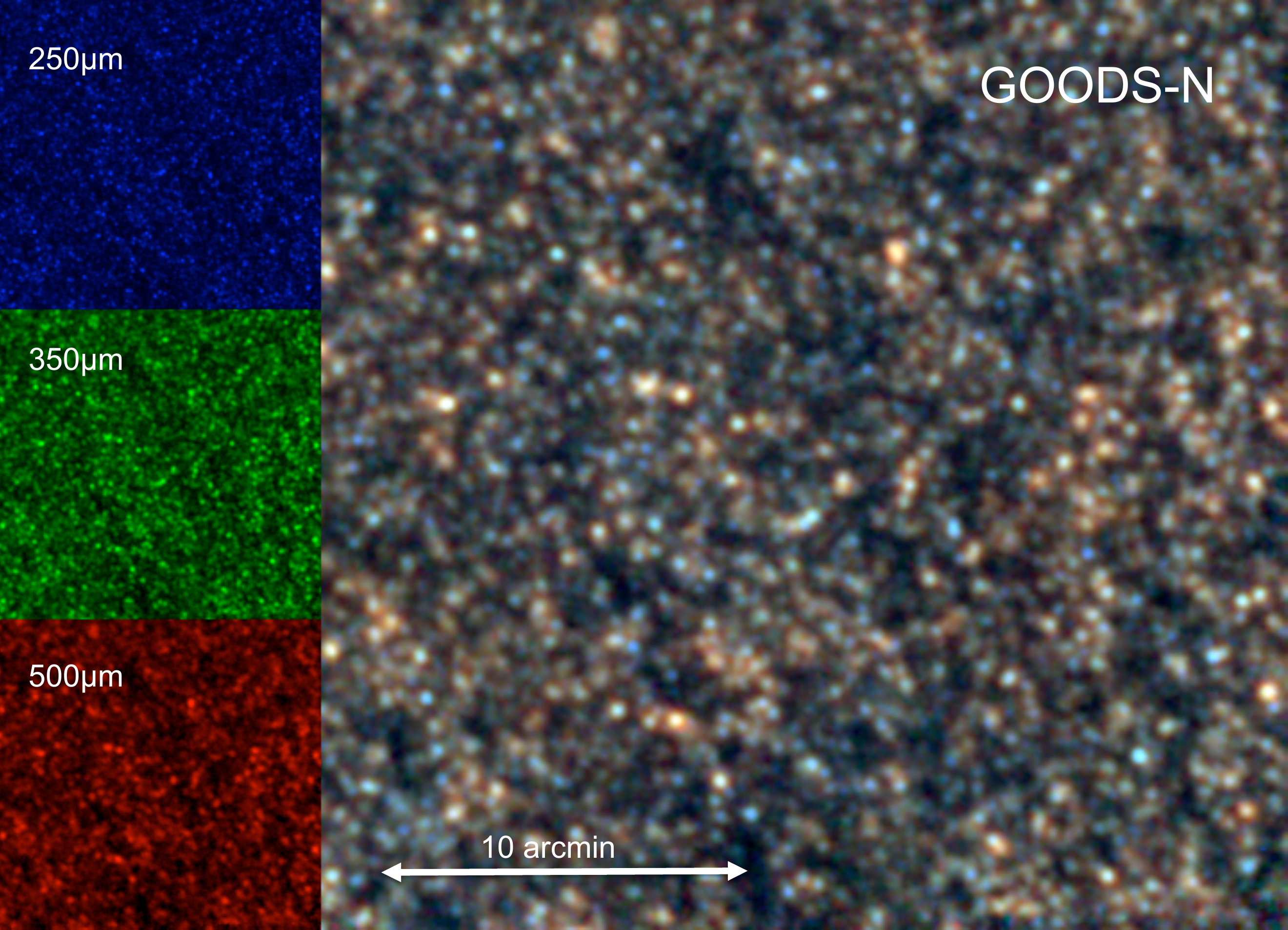}

\caption{Three colour {\em Herschel}-SPIRE image of the GOODS-North region. This is a sub-set of our GOODS-N observation. The left-most panels show the single band images of the cluster, while the central panel shows the three colour image.}\label{fig:goodsn}
\end{figure*}

Our first observation was carried out on 12th September 2009.  This was the first half of our SPIRE observations of Abell~2218 (\#1) and the resulting map from all the data is shown in Fig.~\ref{fig:a2218}.  
This was part of the {\em Herschel} Science Demonstration Phase (SDP). Our SDP observations were designed to exercise most of the modes that were to be used in the full survey, and the SPIRE observations are described in \citet{Oliver2010}.  This includes the observations of GOODS-N (\#14) (Fig.~\ref{fig:goodsn}). The SDP observations concluded on 25th October 2009; AORs are available under the proposal ID  SDP\_soliver\_3.

The program is now being carried out as part of the Routine Phase (proposal ID KPGT\_soliver\_1) and is expected to be completed during 2011. The current ESA schedule is on {\tt herschel.esac.esa.int/observing/ScheduleReport.html} and the observing log can be followed on {\tt herschel.esac.esa.int/observing/LogReport.html}.

\subsection{Comparison with other {\em Herschel} Surveys}
HerMES was planned alongside the `PACS evolutionary Probe (PEP)' survey (Proposal ID KPGT\_dlutz\_1, e.g. \citealt{pep}).  Since then there have been a number of related Key Project surveys carried out in Open Time. 
 There have also {\color{black} been} a few surveys being undertaken in Open Time but not as Key Projects.  These programmes are listed in Table~\ref{tab:herschelprogs}

\begin{table*}
\begin{tabular}{lllrl}
Call & Title & Proposal ID & Time  & Reference\\
       &         &                     &       [hr]         &  \\\hline
Key GT & {\em Herschel} Extragalactic Multi-tiered Survey (HerMES)  & KPGT\_soliver\_1 & 806 & this paper\\
Key GT & PACS evolutionary Probe (PEP)  & KPGT\_dlutz\_1 &  655 & \citealt{pep}\\
Key OT & The Cluster Lensing Survey &  KPOT\_eegami\_1 & 292 &\citealt{Egami2010}\\
Key OT &The {\em Herschel} Astrophysical Terahertz Large Area Survey' (H-ATLAS) &  KPOT\_seales01\_2 &  600 & \citealt{atlas}\\ 
Key OT & The Great Observatories Origins Deep Survey (H-GOODS) & KPOT\_delbaz\_1 & 363 & Elbaz et al.\\
OT1 & The {\em Herschel}-AKARI NEP Deep Survey & OT1\_sserje01\_1 & 74 & Serjeant et al.\\
OT1 & A deep PACS survey of AKARI-Deep field south' & OT1\_ttakagi\_1 & 35 & Takagi et al.\\
OT1 & SPIRE Snapshot Survey of Massive Galaxy Clusters &  OT1\_eegami & 27 & \citealt{Egami2010} \\
OT1 & Measuring the Epoch of Reionization &  OT1\_jcarls01\_3 & 79 & Carlstrom et al.\\
GT2 & HerMES Large Mode Survey  & GT2\_mviero\_1  & 103 & Viero et al. \& this paper\\

\end{tabular}
\caption{{\em Herschel} blank field and cluster lens surveys carried out as Key Programmes or ordinary programmes under Guarenteed Time (GT) or Open Time (OT).}\label{tab:herschelprogs}

\end{table*}

\begin{figure}
\includegraphics[width=8cm]{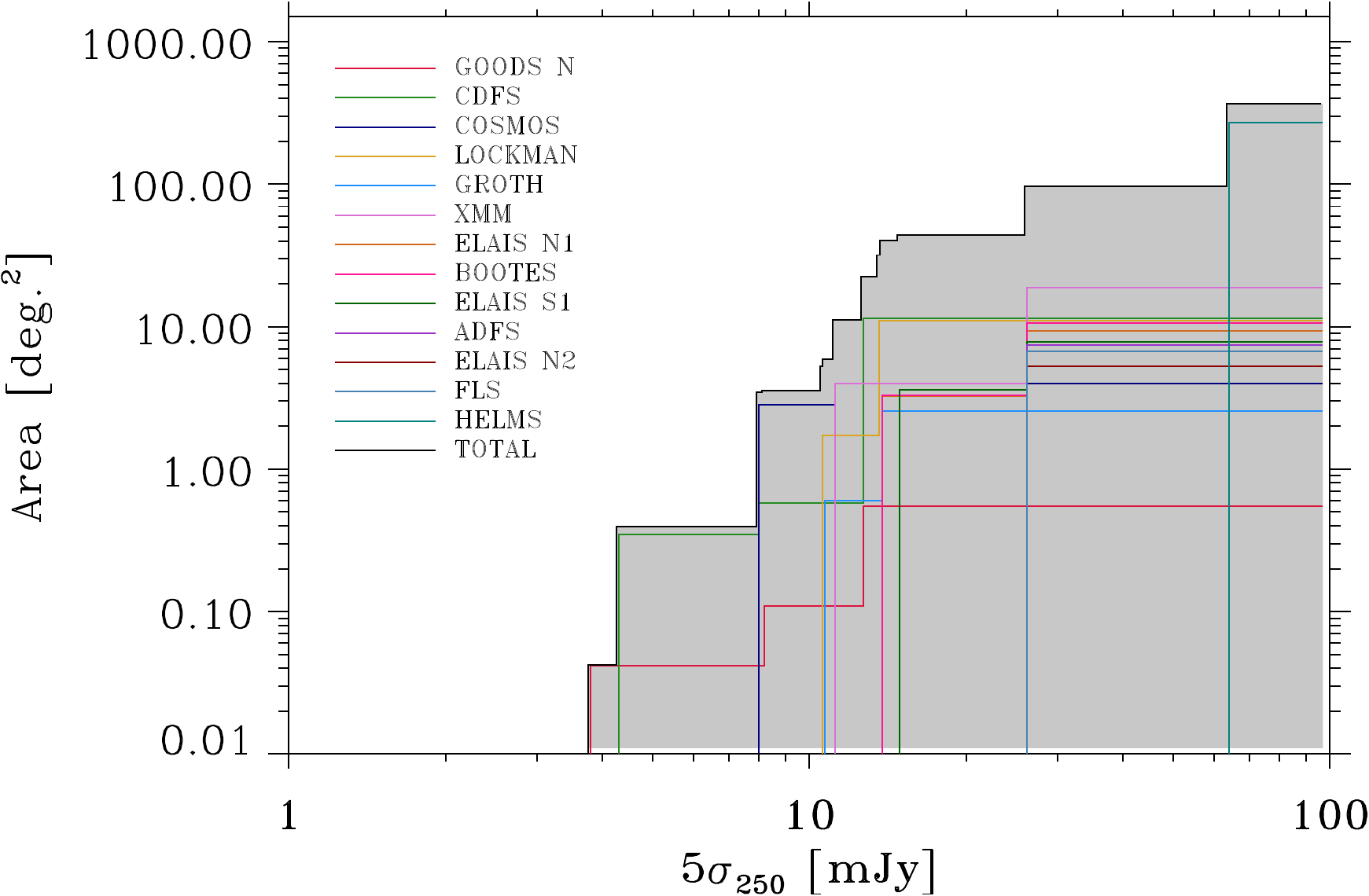}
\caption{Cumulative area against 5-$\sigma$ instrumental noise level at 250$\,$\micron\ for the HerMES blank-field surveys with SPIRE. The colour-coding breaks this down into individual survey fields.}\label{fig:waterfall}
\end{figure}

The cumulative area of all major {\em Herschel}-SPIRE extra-galactic Key Program surveys as a function of instrumental noise (taken from Table~\ref{tab:wcake}) and for the HerMES fields is shown in Fig.~\ref{fig:waterfall}.

\begin{figure}
\includegraphics[width=8cm]{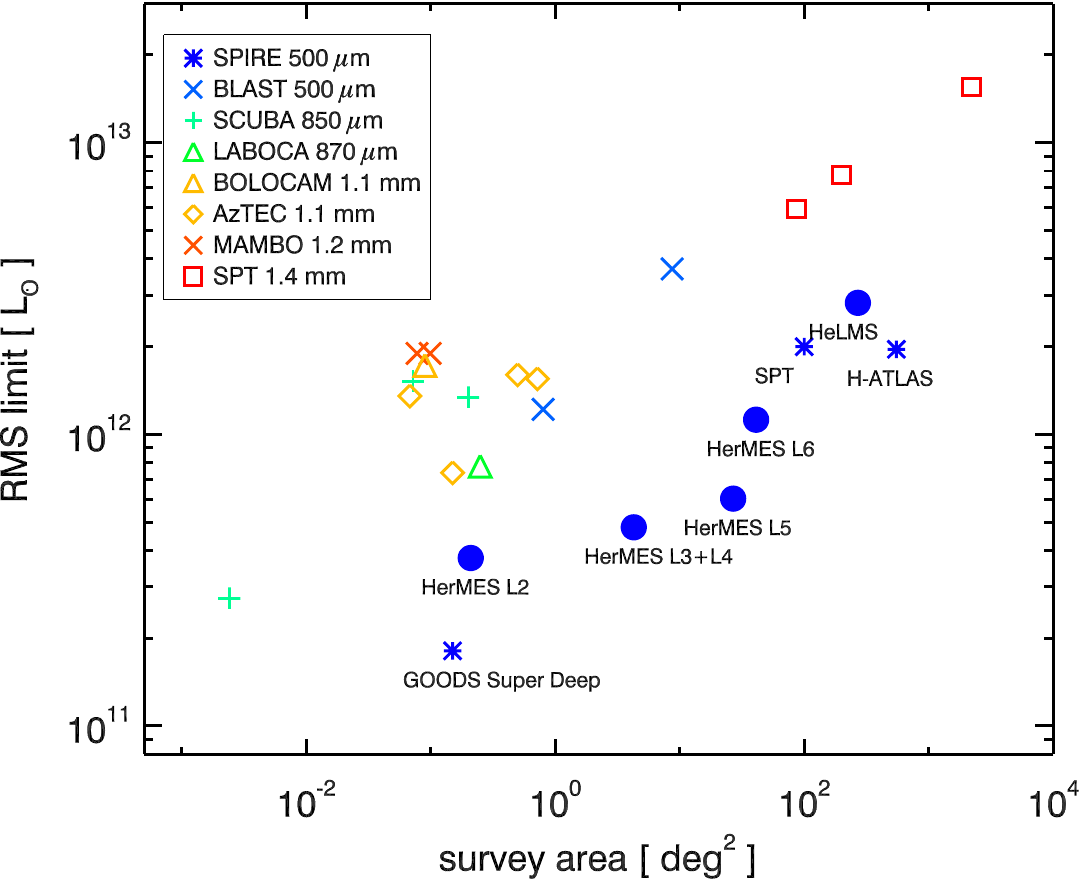}
\caption{Luminosity limit verses redshift for submm surveys to date. The luminosity limit was calculated assuming a modified blackbody of 35K at $z=2$.  
(References for the points are as follows: 
SCUBA -- \protect\citealt{Hughes:1998, Scott2002, Coppin2006},
MAMBO -- \protect\citealt{Greve2004, Bertoldi2007},
BOLOCAM -- \protect\citealt{Laurent2005},
AzTEC --  \protect\citealt{Perera2008, Austermann2010, Scott2010, Aretxaga:2011},
LABOCA -- \protect\citealt{Weiss2009},
SPT -- \protect\citealt{Vieira2010, Williamson2011},
BLAST -- \protect\citealt{blast},
SPT SPIRE -- Carlstrom et al.)}\label{fig:submmsurveys}
\end{figure}

It is striking to compare the {\em Herschel}-SPIRE sub-millimetre surveys with previous sub-millimetre surveys.  To do this we have explored the sensitivity of surveys to a canonical galaxy with a modified blackbody spectral energy distribution with emissivity, $\beta=1.5$, and temperature  $T=35\,$K.  These are shown in Fig.~\ref{fig:submmsurveys}.


\section{Early and Anticipated Science}\label{sec:method}

\subsection{Confusion limits}
An important consideration in design of HerMES was the impact of source confusion at SPIRE wavelengths, i.e. the limited ability to separate individual sources due to the resolution of the telescope and the number density of sources. We define confusion noise to be the standard deviation of the intrinsic variations in a map on the scale of the beam due to all point sources . We planned our survey with reference to several number count models \citep{Lagache:2003, Leborgne2009, Franceschini:2009, Pearson:2009, Xu2003}.  We used these models to estimate the fluctuations in a map which at the 4-$\sigma$ level were 
$1.6\pm0.9, 10.6\pm3.1, 26.3\pm6.3, 32.5\pm7.5$ and $30.0\pm7.5$~mJy at 100, 160, 250, 350 and 500$\,$\micron\ respectively.  The uncertainties come from the scatter between models. The SPIRE confusion noise estimates compare very favourably with the fluctuations in our maps as calculated by \citet{Nguyen2010} with
$5\sigma=24.0, 27.5, 30.5$~mJy  at 250, 350 and 500$\,$\micron, respectively after cutting maps at  $5\sigma$.  This is perhaps fortuitous given that the model counts do not fit the observed counts very well in detail \citep[e.g.][]{Oliver2010, Glenn2010} but may be because the models had been constrained to fit the infrared background.

We planned for the survey to have a substantial area (providing SDSS-like volumes) at the confusion limit, but with some regions well below the confusion limit in very well studied fields, to exploit techniques for mitigating confusion using high signal-to-noise data.

\subsection{Science above the confusion limit}\label{sec:sci1}

\subsubsection{\color{black}Direct determination of the total far infrared luminosity function and its evolution}

Our primary goal has been to determine the total far infrared luminosity function and subsequently the bolometric luminosity of galaxies over the redshift range
$0<z < 3$.  For this analysis we use galaxies detected in {\em Herschel} images combined with extensive multi-wavelength data to
determine photo-$z$s where spectroscopic redshifts are not yet available.

\begin{figure}
\includegraphics[width=8cm]{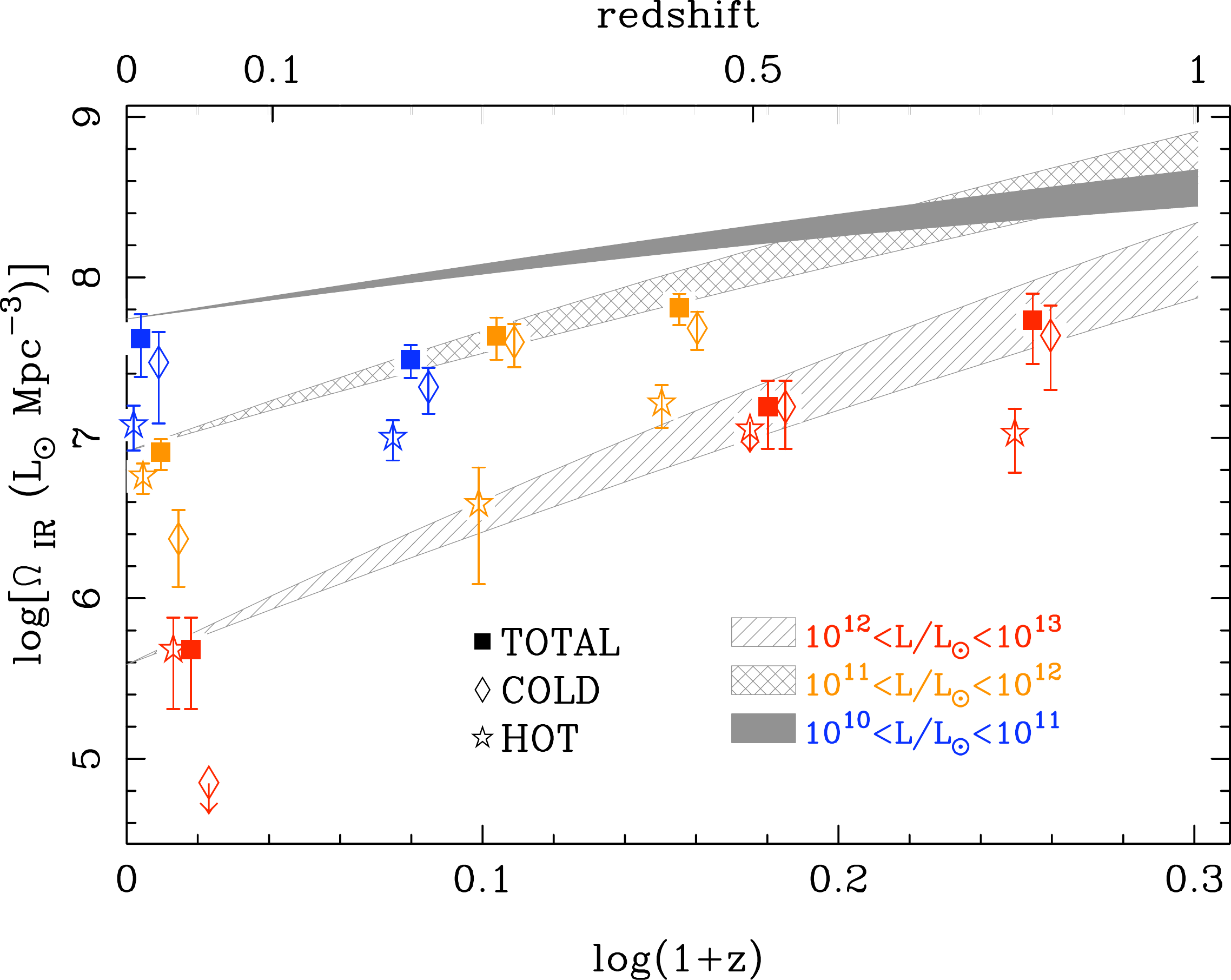}
\caption{
The comoving infrared luminosity density, a proxy for the star
formation history of the Universe, from \protect\citet{Seymour2010}. In grey
shading is the {\em Spitzer} view (following  \protect\citealt{Le-Floch:2005}) showing
the contribution of low IR luminosity galaxies (solid), luminous IR
galaxies (LIRGs, cross-hatched), and rapidly-evolving ultra-luminous
IR galaxies (ULIRGs, hatched), to the total co-moving IR energy
density. Different symbols give subdivision by dust temperature within
each luminosity class.}\label{fig:lefloch}
\end{figure}

Our first results on exploration of the full far infrared SED are given by \citet{Elbaz2010, Rowanrobinson2010, Hwang2010} and \citet{Chapman2010}.  \citet{Elbaz2010} combined photometry from PACS (from the PEP program) and SPIRE (from HerMES). We found that the total far infrared luminosity estimated from extrapolations of {\em Spitzer} 24$\,$\micron\ data agreed well with direct measurements from {\em Herschel} at lower redshift but underestimated the power at higher redshifts (as also seen by \citealt{Nordon2010}).  In that work the longer wavelength (SPIRE) band measurements departed from the model SEDs at lower redshift. This was explored further by \citep{Rowanrobinson2010}, showing that the SPIRE results for some galaxies could be explained with a cold dust component not normally included in canonical templates. Indeed, when simply characterising the SEDs by their effective dust temperature we have shown that the SPIRE detected galaxies cover a broad range of temperatures \citep{Hwang2010, Magdis2010} and thus capture warm objects like the `Optically Faint Radio Galaxies' missed by ground-based sub-millimetre surveys \citep{Chapman2010}.

We have already determined our first measurements of the local luminosity functions  at 250, 350 and 500$\,$\micron\ together with a total infrared (8--1000$\,$\micron) function, finding a local luminosity density of $1.3^{+0.2}_{-0.2}\times10^8{\rm L}_{\odot}\, {\rm Mpc}^{-3}$ \citep{Vaccari2010} and 
showing that the 250$\,$\micron\ function evolves strongly to $z\sim 1$ \citep{Eales2010}, similarly to earlier studies at shorter wavelengths.
Future analysis (in preparation) will study wider areas with more and better ancillary data and extend these results to higher luminosities, higher redshifts and model the relative contribution of AGN and star-formation to the bolometric emission, as well as exploring the relation between the infrared luminosities and the stellar properties probed at optical, NIR and UV wavelengths.

\subsubsection{\color{black} Star-formation and environment}

Environment on various scales plays an important role in the process of galaxy formation.  Perhaps the most striking observational evidence is that clusters today have a much higher fraction of early-type galaxies than is found in the field. Likewise the successful physical models of galaxy formation predict a very strong co-evolution between galaxies and dark-matter halos.

There are many ways of determining the role of environment observationally:  one can directly 
examine the galaxy properties (e.g. the SFR distribution functions) in different 
environments; one can explore the environments
of galaxies in different luminosity classes; one can use the clustering of particular galaxy populations
to infer the mass of the dark matter halos in which they are located,
to relate these to their present-day descendants; or one can directly use the structure in the maps to constrain such models.  All these methods have the same basic requirement, a volume sufficiently
large to sample enough of the environments of interest, and sufficiently deep to constrain the populations of interest.
A simulation in Fig.~\ref{fig:r0} shows that we could discriminate different halo mass hosts for different sub-classes of galaxies and compare the clustering of the FIR galaxies with quasars from optical studies.  

\begin{figure}
\includegraphics[width=8.cm]{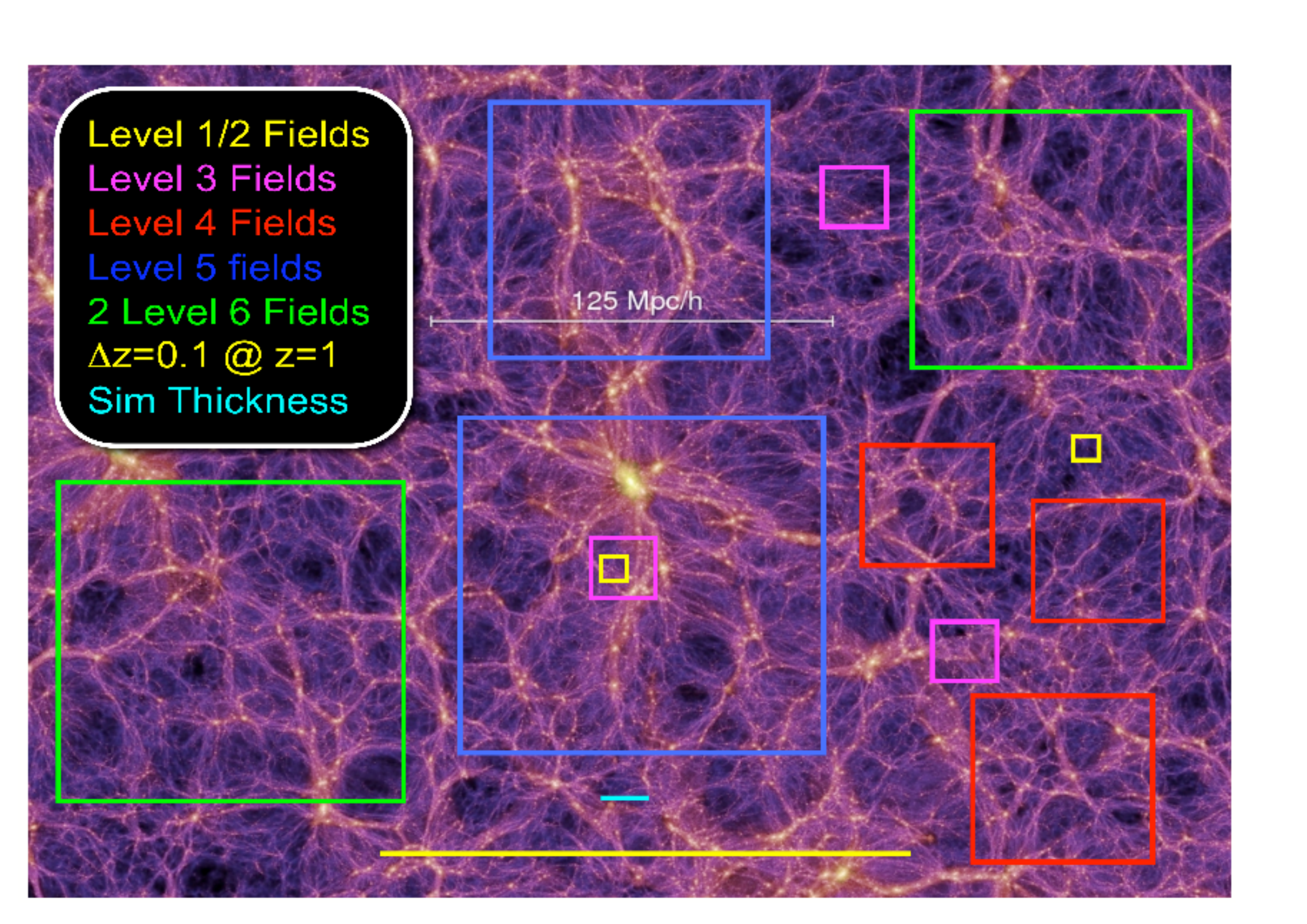}
\caption{A slice of the dark matter in the Millennium Simulation of the Universe, seen today \citep{Springel:2006}.
Overlayed are the footprints of some of our fields, showing how much of this slice they would sample at $z=1$. {\color{black} This thin slice exaggerates the effect but illustrates that to} overcome sampling variance and to probe a full range of environments we need multiple, large fields.}\label{fig:virgo}
\end{figure}

\begin{figure}
\includegraphics[width=8.cm]{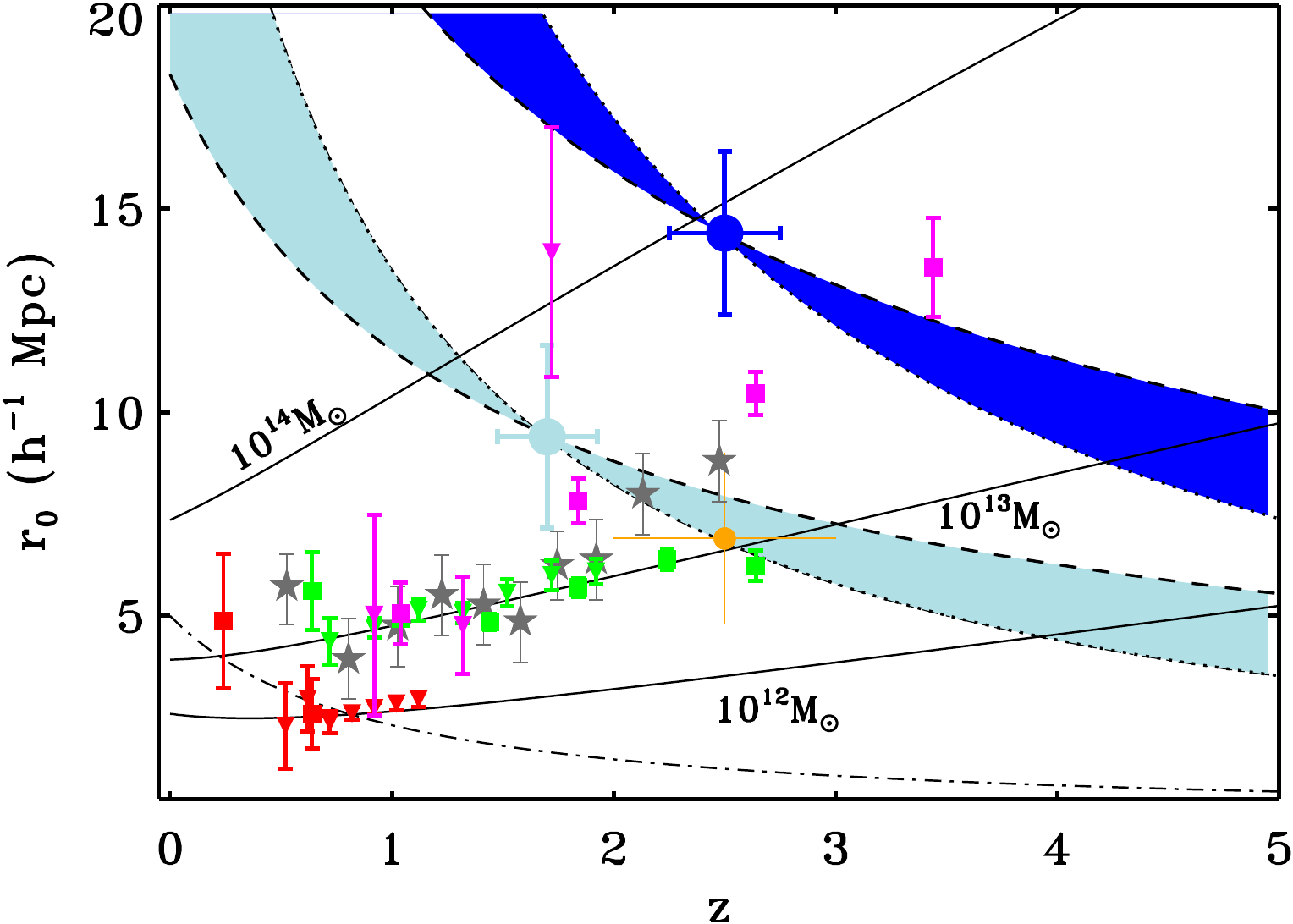}
\caption{Evolution of
co-moving correlation length, $r_0$, with redshift.  Solid lines show the predicted clustering amplitude of haloes of given
mass. We have simulated data for the clustering of LIRGs (red), ULIRGs (green) and HLIRGs (magenta), assuming they inhabit
halos of mass  $10^{12}, 10^{13}$ and $10^{13.5} {\rm M}_\odot$ respectively.  The simulation is for our 250$\,$\micron\ surveys at Level~5
(square) and Level~6 (triangle). For comparison we show quasar clustering from \citet{Croom:2005} as stars and SCUBA galaxies from \citet{Blain:2004} as orange circles. {\em Spitzer} sources from \citet{Farrah:2006} are shown as blue circles and blue shaded regions extrapolate those objects to their progenitors and descendants.}\label{fig:r0}
\end{figure}

First results on the clustering of HerMES galaxies were given by \citet{Cooray2010}, indicating that the HerMES sources with $S_{250}>30\,{\rm mJy}$ (at $z\sim2$) were in dark matter halos with masses above $(5\pm4)\times 10^{12}\,{\rm M}_\odot$.  

Clustering can also be used in other ways. A recent cross-correlation analysis indicates that there is a correlation between HerMES sources at $z\sim 2$ and foreground galaxies from SDSS at $z\sim 0.2$ and SWIRE at $z\sim0.4$ \citep{Wang2011}. While some of this signal can be attributed to the intrinsic correlation of galaxies in the overlapping tails of the redshift distributions, there is clear evidence for a signal arising from the amplification of the HerMES source fluxes by lensing from foreground galaxies.

\subsubsection{\color{black}Extreme galaxies}
Rare objects provide challenges for theories and may expose important but transitory phases in the life-cycle of galaxies. The
very wide surveys, in particular, will discover many exotic objects, which are prime targets for  ALMA. Galaxies with extremely high star formation rates would be difficult to explain with some models of galaxy formation.   Limited area sub-millimetre surveys have already discovered small samples of galaxies with very high star formation rates $\ga1000 {\rm M}_{\sun}{\rm yr}^{-1}$ e.g. SMM~J02399-0136 \citep{Ivison:1998}, GN20 \citep{Daddi:2009,Borys:2003} and 
 MIPS~J142824.0+352619
\citep{Borys:2006}. By mapping large areas at the wavelengths where re-emission from star formation peaks, we will be able to quantify the number density of systems of $\sim1000 {\rm M}_\odot{\rm yr}^{-1}$ and determine whether there are any systems with even higher star formation rates.  Even individual examples of such systems would be important as extreme astrophysical laboratories and would provide fruitful targets for new facilities, especially  ALMA.

A primary search tool will be the SPIRE colours.  Searches have already revealed many galaxies \citep{Schulz2010} with very red colours $S_{250}/S_{350} < 0.8$ and with flux densities above 50~mJy. These  may be a mix of intrinsically cool galaxies at lower redshift, and galaxies at high redshift, including some that are lensed by foreground galaxies.

\subsubsection{\color{black}Lensed Systems}

Lensed systems are interesting because, although lensing is a rare phenomenon, they provide a magnified view of more common, relatively normal, but distant galaxies, which can then be easily studied.  
An example of a lensed source found in early HerMES data is
HERMES J105751.1+573027, a $z=2.957$ galaxy multiply lensed
by a foreground group of galaxies.  Coupled with a lensing model
derived from high-resolution observations \citep{Gavazzi2011}, 
the magnification and large image separation allowed us to
investigate the continuum SED from the optical to far-IR \citep{Conley2011}, as well as model the CO line excitation \citep{Scott2011}
and study the gas dynamics \citep{Riechers2011}.

\subsection{Science below the confusion limit}\label{sec:sci2}
The deepest observations at SPIRE wavelengths suffer substantial confusion noise due to faint unresolved galaxies, and are limited in their ability to define true luminosities, SEDs and physical conditions
within the most active galaxies during the peak epoch of galaxy formation at redshift $z \sim 2$.
We will investigate and employ super-resolution techniques, e.g. {\sc CLEAN} \citep{clean} or matched filtering \citep{Chapin2011}.  However, as argued in \citet{Oliver2001}, we expect the gains from  blind image deconvolution techniques to be modest except at the very highest signal-to-noise ratios.

One approach to combat the problem is to study isolated sources as we have discussed in \citet{Elbaz2010, Brisbin2010} and \citet{Schulz2010}, however, we are pursuing many other mitigating techniques. 

\subsubsection{\color{black} Ultra-deep far-infrared galaxy surveys from imaging of rich clusters of galaxies} 
Rich clusters can be used as tools to
mitigate this effect, allowing high-redshift galaxy formation to be investigated by the
gravitational magnification of the primordial galaxies behind the cluster. This has been demonstrated at relevant wavelengths by \citet{Smail:2002}, \citet{Cowie:2002}, \citet{Metcalfe:2003}, \citet{Chary:2005} and \citet{Swinbank:2010}. 

Gravitational lensing brightens and separates the images of all background galaxies
within 1--2\arcmin\ of the core of the cluster \citep[e.g.][]{Kneib:2004}, making individual background galaxies
easier to detect.  This also allows the sources of up to about 50 per cent of the
otherwise confused and unresolved background radiation to be identified 
with specific galaxies.

The selected clusters have some of the best archival data
available, including deep {\em HST} ACS/NICMOS images, 
ultradeep $\mu$Jy radio imaging, deep mid-IR imaging from {\em Spitzer}, and 
X-ray images from {\em Chandra}/{\em XMM-Newton}. The mass and magnification profiles are known accurately, 
from extensive spectroscopy of multiply-lensed images \citep{Kneib:1993}.

Our observations of 10 clusters will provide about 180 sources that will allow us to quantify the space density of the faintest
{\it Herschel} galaxies with 10 per cent accuracy.   Two clusters (Abell~2218 and Abell~1689) were believed, in advance, to be relatively free of bright lensed galaxies.  This was intentional as these were originally intended for very deep observations in order to detect of order 10 even fainter lensed sources (to
determine the counts of {\it Herschel} galaxies at the 5-mJy detection level, reaching well below the blank-field confusion limit) so we wanted to avoid confusion from known lensed galaxies. Following modification to our program in the light of  analysis of in-flight data  only Abell~2218 was observed deeper than the others.  
The results from our SDP cluster observation of Abell~2218 clearly demonstrate that we can detect high redshift lensed galaxies, see Fig.~\ref{fig:a2218}.

\subsubsection{\color{black} Multi-colour 1-point fluctuation analysis below the confusion limit}
\begin{figure}
\includegraphics[width=7.7cm]{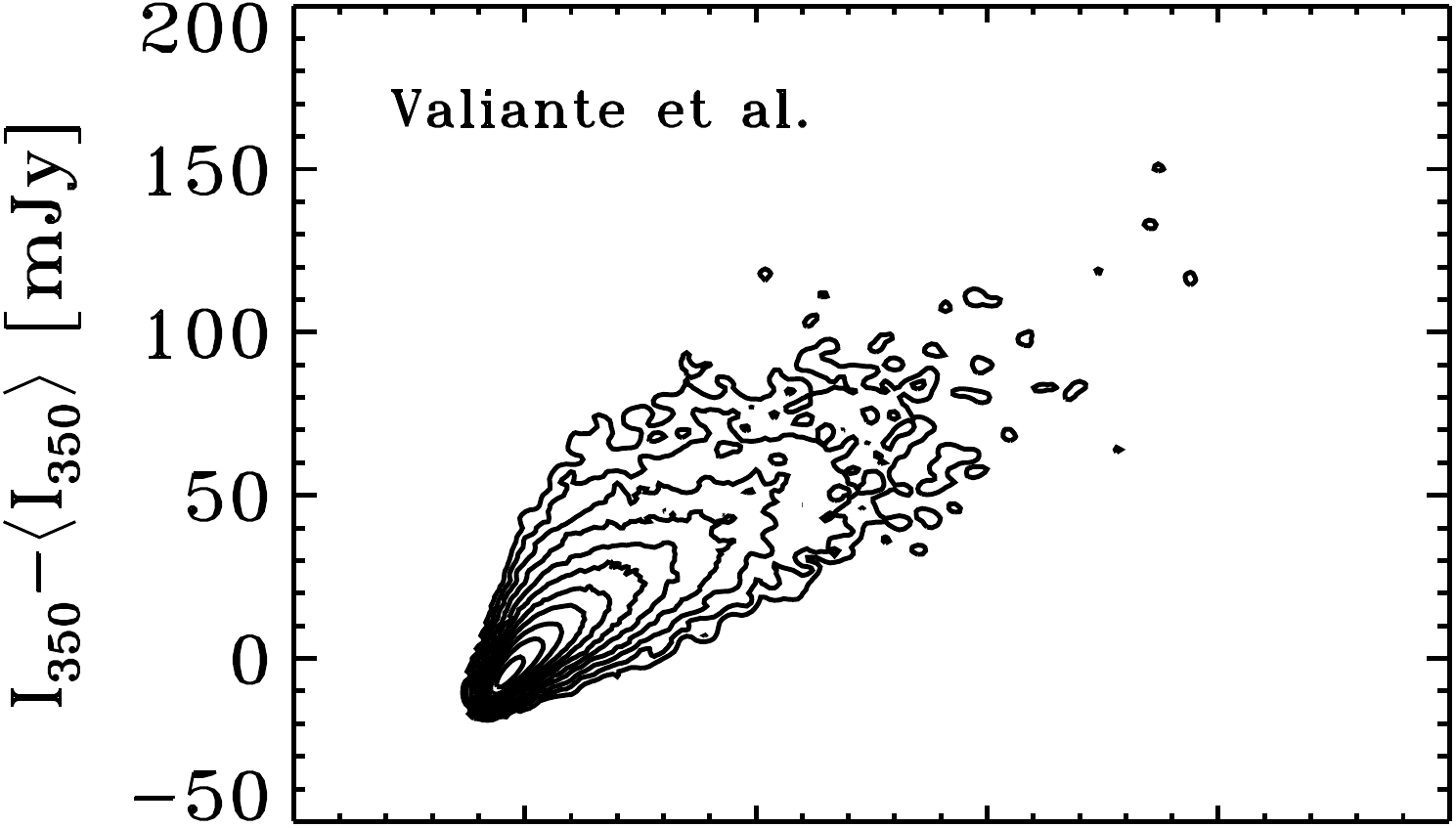}
\includegraphics[width=8cm]{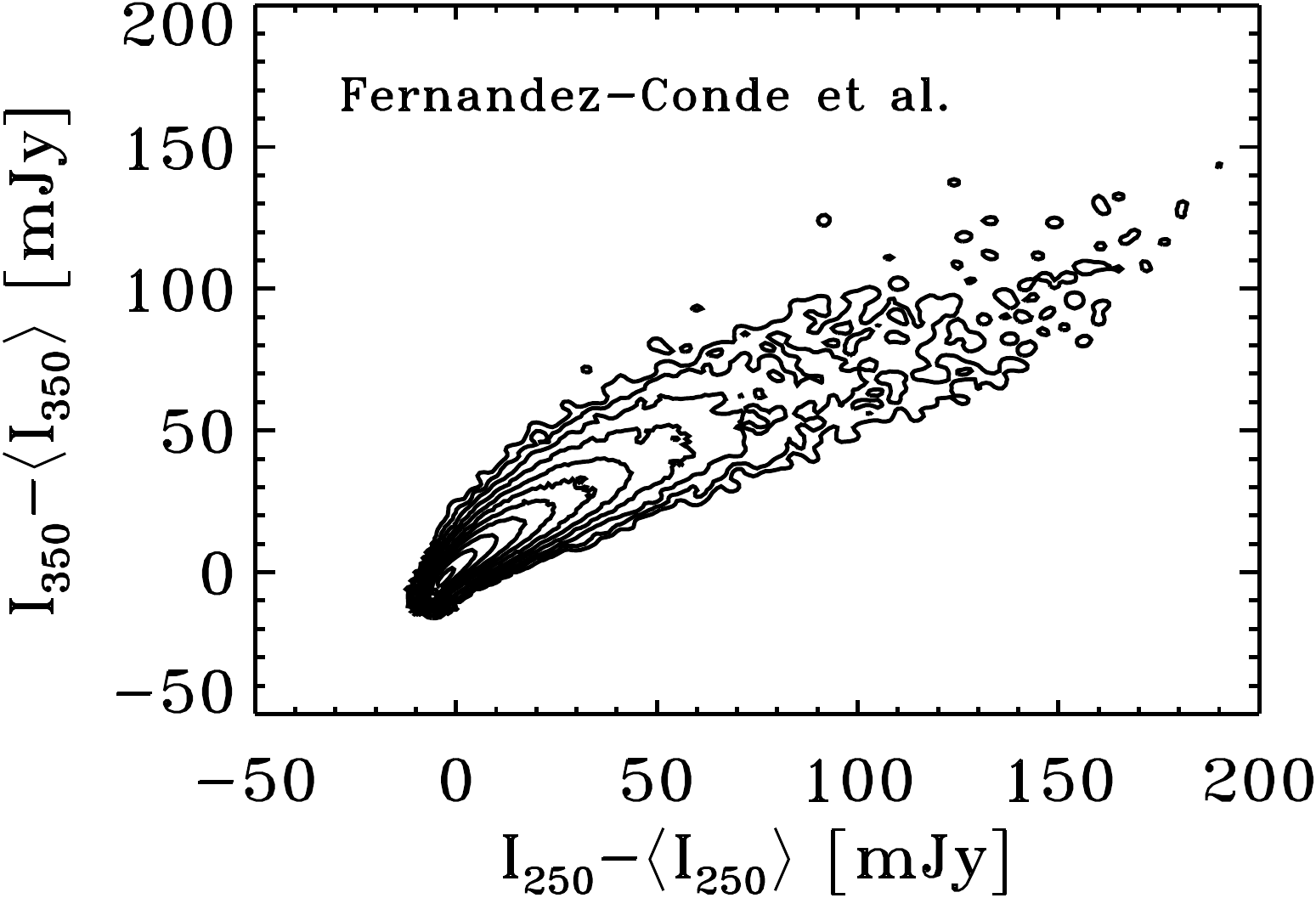}
\caption{Simulation of a two dimensional $P(D)$ analysis, showing discrimination between models. The $x$ and $y$ axes show the pixel intensities (in mJy beam$^{-1}$) in the 250$\,$\micron\ and 350$\,$\micron\  bands, respectively. The contours show the number of pixels with those intensities, logarithmically spaced. The top panel is for the number count model of \citet{Valiante2009}, the bottom is for the mock catalogues  of \citet{FernandezConde2008} based on the models of \citet{Lagache:2003}. The simulations are around ten deg$^2$ and with 1$\,$mJy of Gaussian noise in each band.} \label{fig:pofd}
\end{figure}

Analysis of the fluctuations in the cosmic IR background radiation provides unique information on 
sources too faint to be detected individually \citep[see, e.g.][]{Maloney:2005, Patanchon:2009}. Our Level~2 and Level~3 fields allow us to analyze the fluctuation distribution down to flux densities of  2--3~mJy, where much of the background was expected to be resolved.  By analyzing the fluctuations in all three SPIRE wavebands, we can obtain statistical information on SEDs. This multi-colour $P(D)$ analysis provides a powerful method for distinguishing different number count models, thereby constraining the redshifts and emission properties of the source population (Fig.~\ref{fig:pofd}).  This requires very precise characterization of the instrument noise for optimal analysis.

We undertook a mono-chromatic fluctuation analysis using three fields from our SDP data. With that analysis \citep{Glenn2010} we reached a depth of  2 mJy$\,{\rm beam}^{-1}$, significantly deeper than any previous analysis at these wavelengths.  Modelling this distribution with parameterised number counts confirmed the results from resolved sources \citep{Oliver2010} and was in disagreement with previous models. The fits accounted for 64, 60, and 43 per cent of the far-infrared background at 250, 350 and 500$\,$\micron, respectively.

\subsubsection{\color{black} Average SEDs of galaxies contributing to the infrared background}
Prior information from shorter wavelength (e.g., 24$\,$\micron\ with MIPS) can be used to infer the statistical properties (such as source density or SEDs) at longer wavelengths.  
A more promising route to achieving super-resolution results is to use prior information on the positions of sources from higher resolution data at other wavelengths.  This has been demonstrated with HerMES data in \citet{Roseboom2010} acheiving robust results for source fluxes down to $S_{\rm 250}\approx 10\,{\rm mJy}$. 

\begin{figure}
\includegraphics[width=8cm, height=8cm]{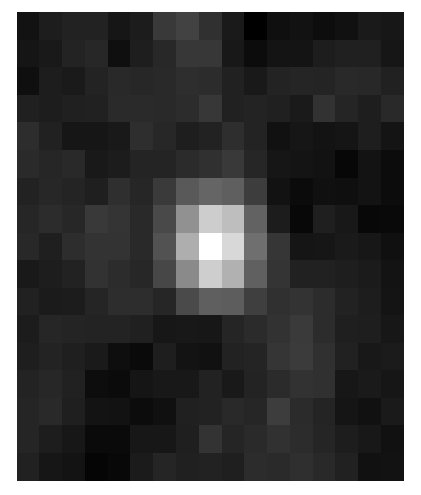}
\caption{Example of about 4000  {\em Spitzer} IRAC selected `Bump-3' sources (i.e., objects with peak emission at 5.8$\,$\micron) stacked in HerMES SPIRE maps at 250$\,$\micron\ with 6\arcsec\ pixels. The clear detection allows one to derive aggregate SEDs of this galaxy population, expected to lie at $2.2<z<2.8$.}\label{fig:stack}
\end{figure}

A related technique is `stacking', which averages the signal from many similar prior sources.  In the absence of significant correlations the confusion variance would then reduce in proportion to the number of prior sources in the `stack'.
Stacking has been successfully applied to {\em Spitzer} MIPS data; \citet{Dole:2006} stacked more than 19,000 24$\,$\micron\ galaxies to find the
contributions of the mid-IR galaxies to the far-IR background (70 and
160$\,$\micron). With this technique, they gained up to one order of
magnitude in depth in the far-IR. It appears that a large fraction of
the 24$\,$\micron\ sources can be statistically detected at longer
wavelengths \citep[e.g.][]{Marsden:2009}. Such an analysis applied to {\em Herschel} will allow us to
extend galaxy SEDs to the FIR/sub-mm to quantify the
contribution of different populations to the background \citep[e.g.][]{Dye:2007, Wang:2006}, or to explore the star-formation properties as a function of redshift and stellar mass  \citep[e.g.][]{Oliver:2010a}. Such procedures might use {\em Spitzer} 24$\,$\micron\ catalogues and/or the PACS catalogue.  This type of analysis is critically dependent on the quality and depth of the ancillary data, and further motivates our choice of very well studied extra-galactic fields. An example of this approach is shown in Fig.~\ref{fig:stack}.

Stacking has already been used in some of our analysis \citep[e.g.][]{Ivison2010, Rigopoulou2010} and our first results analysing the contribution of various prior populations to the background through stacking will be presented by  \citet{Vieira2010b}.

\subsubsection{\color{black}Extragalactic Correlations Fluctuations}

A comprehensive fluctuations analysis is an essential complement to the aspects of our survey allowing us to investigate the majority population of objects, those below the {\em Herschel}
confusion limit.  Using the two shallowest tiers of the survey, we can specifically target non-linear clustering on angular scales $< 10\arcmin$, virtually inaccessible to {\em Planck}, and where SPIRE is not susceptible to {\color{black} low frequency drifts}.  
The clustering of undetected sources produces fluctuations on larger spatial scales \citep{Amblard:2007,Haiman:2000,Knox:2001} which are expected to be brighter \citep{Scott:1999} than Poisson fluctuations on spatial scales $> 1$ \arcmin.  On large angular scales, background fluctuations measure the linear clustering bias of infrared galaxies in dark matter halos.  On small angular scales, fluctuations measure the non-linear clustering within individual dark matter halos, and the physics governing how FIR galaxies form within a halo as captured by the occupation number of FIR sources.  
This halo approach \citep[e.g.][]{Cooray:2002} will allow us to compare the results of a {\em Herschel} fluctuations survey with studies at other wavelengths, to obtain a consistent picture of galaxy clustering and evolution.  
Finally, this fluctuation survey is designed to complement surveys by {\em Planck}  on larger angular scales.   

\begin{figure}
\includegraphics[width=8cm]{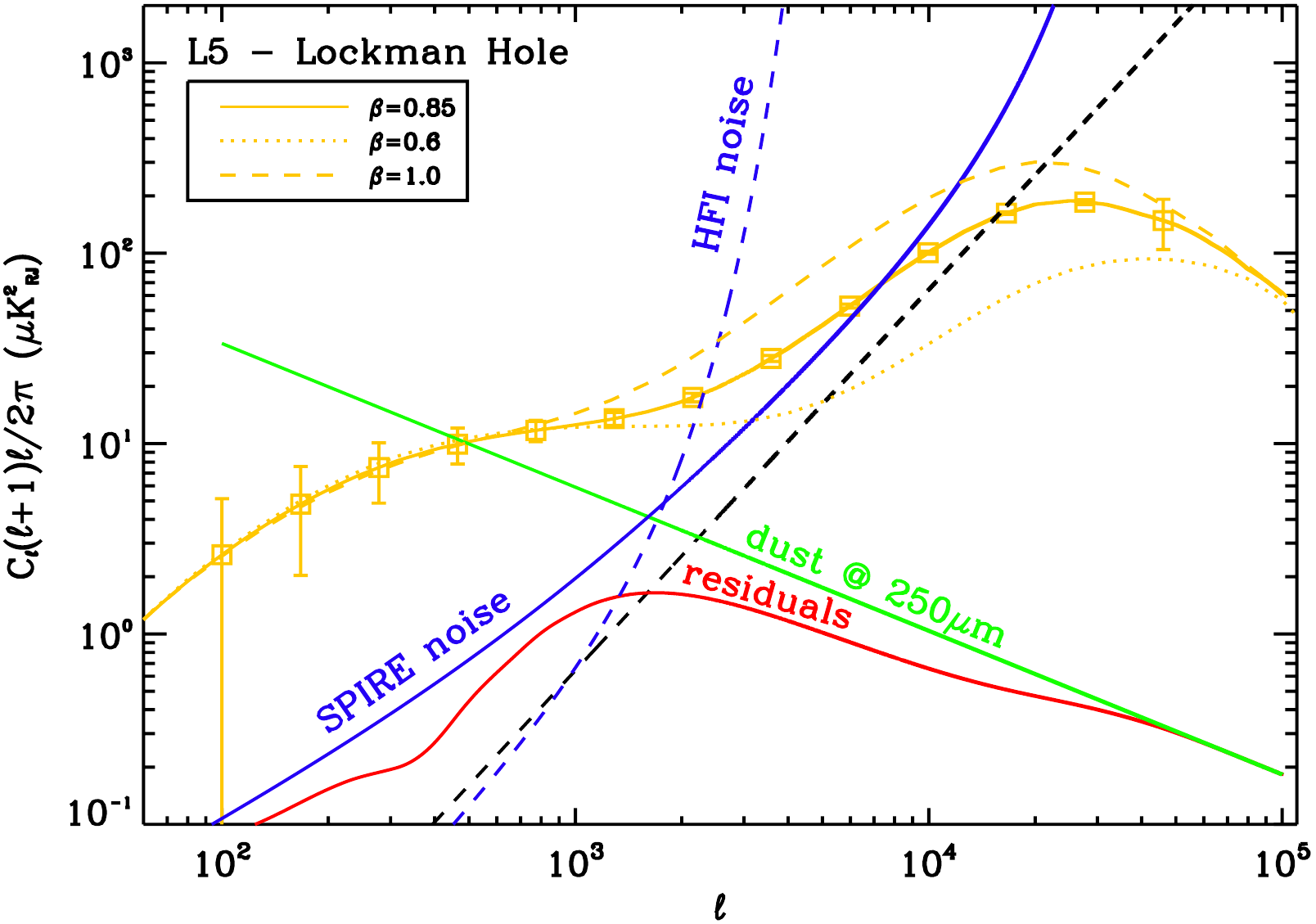}
\caption{The angular power spectrum of unresolved anisotropies at 350$\,$\micron.  We model the power spectrum under the halo approach and describe non-linear clustering with a halo occupation number $\beta$, as shown by the orange curves. We show simulated binned errors with SPIRE in the 11 deg$^2$ Lockman Hole L5 field, including both instrument noise and sample variance, and removing shot noise from galaxies below the detection limit (dashed black curve).  For reference, the long-dashed and solid blue lines show the noise per multipole for {\em Planck} and SPIRE, respectively. The green line
is the foreground dust spectrum, determined for the same field using dust maps.  In red we show the residual foreground spectrum after cleaning with multi-wavelength data.  Even if not removed,
dust does not contaminate small angular scales, where SPIRE excels.}\label{fig:cl}
\end{figure}

First measurements of correlated fluctuations from clustered infrared galaxies at sub-mm wavelengths have been detected by \citep{Lagache2007, grossan2007,viero2009,hall2010,dunkley2011,hajian2011}.  
Our first results \citep{Amblard2011} have extended these findings at arcminute scales by measuring the non-linear 1-halo component for the first time.  Modelling suggests that at 350$\,$\micron\ 90 per cent of the  
background intensity is generated by faint galaxies at $z > 1$ in dark matter halos with a minimum mass of $\log[M_{\rm min}/{\rm M}_\odot] = 11.5^{+0.7}_{-0.2}$, in agreement with BLAST \citep{viero2009}.

\subsection{\color{black}Additional Science Enabled by HerMES}

We expect to detect over 100,$\,$000 sources in our survey. The scientific themes explored in sections~\ref{sec:sci1} and \ref{sec:sci2} will be dramatically extended and improved with the samples available now and the full sample once complete.  Here we mention briefly a very few other science topics that might be addressed by us or others using such a large survey.

The FIR colours of the {\em Herschel} sources can help addressing the question of how much of the energy production comes from accretion (AGN) and how much from star formation. First results on an SDSS sample of AGN \citep{Hatz2010} find that one third are detected by SPIRE, with the long wavelength colours indistinguishable from star forming galaxies. Modelling of the full SED required the combined contribution of both AGN and starburst components, with the former dominating the emission at the MIR wavelengths and the latter contributing mostly to the FIR wavelengths. This suggests that SPIRE detects the star formation in AGN, with little contamination from any dusty torus, offering high hopes for disentangling nuclear and star formation activity.

The wealth of data in these fields mean we can explore the FIR properties of many known samples.  Our first results on Lyman break galaxies have already shown that we can detect U-band dropout sources with stacking \citep{Rigopoulou2010} and FUV drop-out sources individually \citep{Burgarella2011}.  We have also shown that galaxies selected on the basis of the {\em Spitzer} IRAC colours probe a wide range of FIR temperatures \citep{Magdis2010}.

We will compare the FIR measure of star-formation with other tracers.  In collaboration with the PEP team we examined the well-known FIR radio correlation in GOODS-N \citep{Ivison2010}.  Exploring $q_{\rm IR}$, i.e. the  logarithmic ratio of the rest-frame 8--1000$\,$\micron\ flux and the 1.4-GHz flux density, there is no  evidence that $q_{\rm IR}$ changes signiÞcantly for the whole sample: 
$q_{\rm IR} \propto (1 + z) ^\gamma$, where $\gamma=-0.04 \pm 0.03$ at $z = 0-2$, although if the small volume at $ z < 0.5$ is removed 
 we find $\gamma=-0.26 \pm 0.07$. 
 HerMES will create a complete data set to understand the global relationship between FIR and optical galaxies, the effect of dust attenuation in optical/UV populations, and phenomena in individual galaxies.  First results comparing HerMES and {\em GALEX} \citep{Buat2010} confirm that total infrared luminosity accounts for 90 per cent of the total star formation rate, though this reduces to 70 per cent when considering the lower star formation rate systems ($\dot{M_*}<1 {\rm M}_{\sun}{\rm yr}^{-1}$).
 
These ancillary data can also be used to investigate the detailed properties of the FIR galaxies, e.g. their morphology. One study has explored galaxies with morphological classifications at $2<z<3$ and shows that the mean SFR for the spheroidal galaxies is about a factor of three lower than for the disk like galaxies \citep{Cava2010}.

Observations of the rich clusters -- the densest known regions of the Universe -- yield information about their astrophysics
and history via the Sunyaev-Zel'dovic (SZ) effect \citep{Birkinshaw1999,Carlstrom:2002}, which dominates the extended several-arcmin-scale
emission of clusters at wavelengths longer than about 500$\,$\micron.  The SZ effect arises from inverse Compton scattering of cosmic microwave background photons by hot (1--10\,keV) gas in the intracluster medium. We intend to combine SPIRE and {\em Planck} data to
measure the SZ effect and the sub-millimetre foregrounds between 150\,GHz and 1\,THz. Based on the different spectral shapes of the SZ
effect and foregrounds, SPIRE data will enable us to separate out Galactic dust, cluster and background galaxies, the thermal
SZ effect and the effects of relativistic electrons.

\section{Data Products}\label{Sec:Data}
\subsection{SPIRE catalogues}
As an illustration of the kind of data products that HerMES will produce we show an approximation to the SPIRE 250$\,$\micron\ survey areas and depths in Table~\ref{tab:250counts} (together with H-ATLAS and GOODS-H).  We indicate an estimate of the numbers of galaxies on the sky from the \citet{Valiante2009} model which is one of the best fits to the current data and to a direct determination of the counts from both resolved sources \citep{Oliver2010} and fluctuation analyses \citep{Glenn2010}. Finally we give an estimate of the numbers of catalogued sources above those flux density limits estimated from our 24 \micron\ driven extractions (at deep levels) and our single-band detections at shallow levels.  Overall we thus expect 100,000 sources detected at $>5 \sigma$.   

\begin{table}
\begin{tabular}{lrrrrrr}
Levels & Area & \multicolumn{1}{c}{$5\sigma_{\rm 250}$} & $N_{\rm  Val.}$  & $N_{\rm Glenn}$ & $N_{\rm cat}$\\
& [deg$^2$] &  [mJy] & [$10^3$]  & [$10^3$] & [$10^3$] \\\hline
PACS Ul. & 0.012 & \\
Level 1        & 0.15 & 4 &  2.2   & $2.0\pm0.1$ & ---\\
Levels 2-4     & 6.0 & 10 & 17 &  $22.4\pm0.9$\\
Level 5        & 37    &15 & 53 & $73.6\pm2.3$ &52\\
Level 6        & 52    & 26 & 20  & $28.1\pm0.6$ &30\\
H-ATLAS & 570 & 45 & 76 & $90.6\pm2.9$&115\\
Level 7 (HeLMS)        & 270 & 64 & 130 &  & 24\\


\end{tabular}
\caption{Projected SPIRE survey results for the 250$\,$\micron\ band.  This table simplifies the survey giving approximate instrumental noises in 4 tiers (L1 includes GOODS-N).  The $5\sigma$ confusion noise from \protect\cite{Nguyen2010} is  29~mJy, approximately the Level~6 depth. Numbers of 250$\,$\micron\ sources are {\color{black} estimated from: a count model \protect\citep[$N{\rm val}$]{Valiante2009}; our $P(D)$ analysis \protect\citep[$N_{\rm Glenn}$]{Glenn2010} and from our raw number counts in fields that we have at these depths, extracted as described in 
\protect\citep[$N_{\rm cat}$]{Smith2010}}. }
\label{tab:250counts}
\end{table}

\subsection{Ancillary Data}\label{sec:ancillary}

\subsubsection{Required Ancillary data}
To estimate the required ancillary data we have examined our first cross-indentified catalogues \citep{Roseboom2010}.  These are lists with photometry at the positions of known 
24$\,$\micron\ galaxies and thus are not a complete description of the {\em Herschel} populations; however they are approximately 90 per cent complete. 

\begin{figure}
\includegraphics[width=8cm]{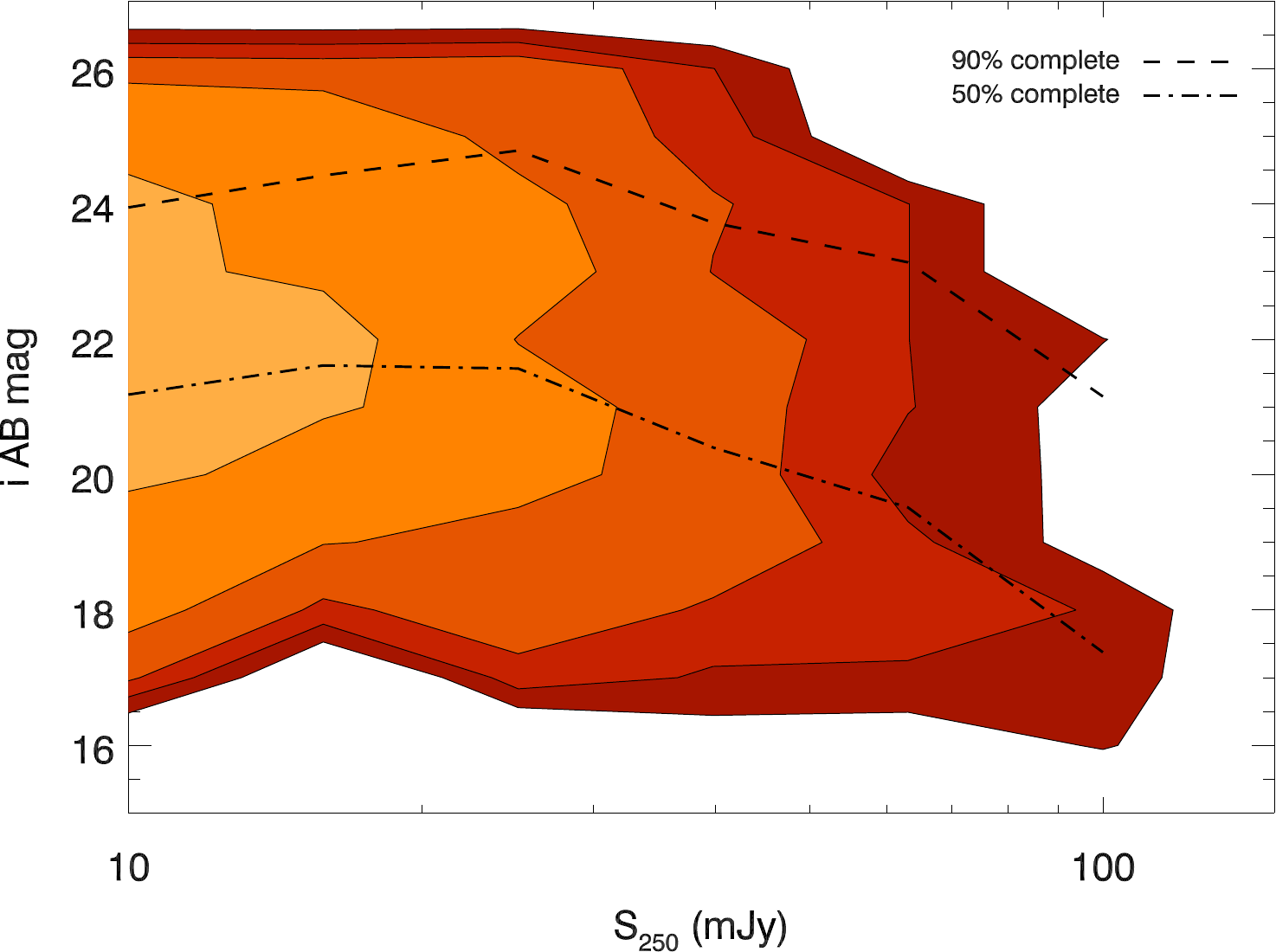}
\includegraphics[width=8cm]{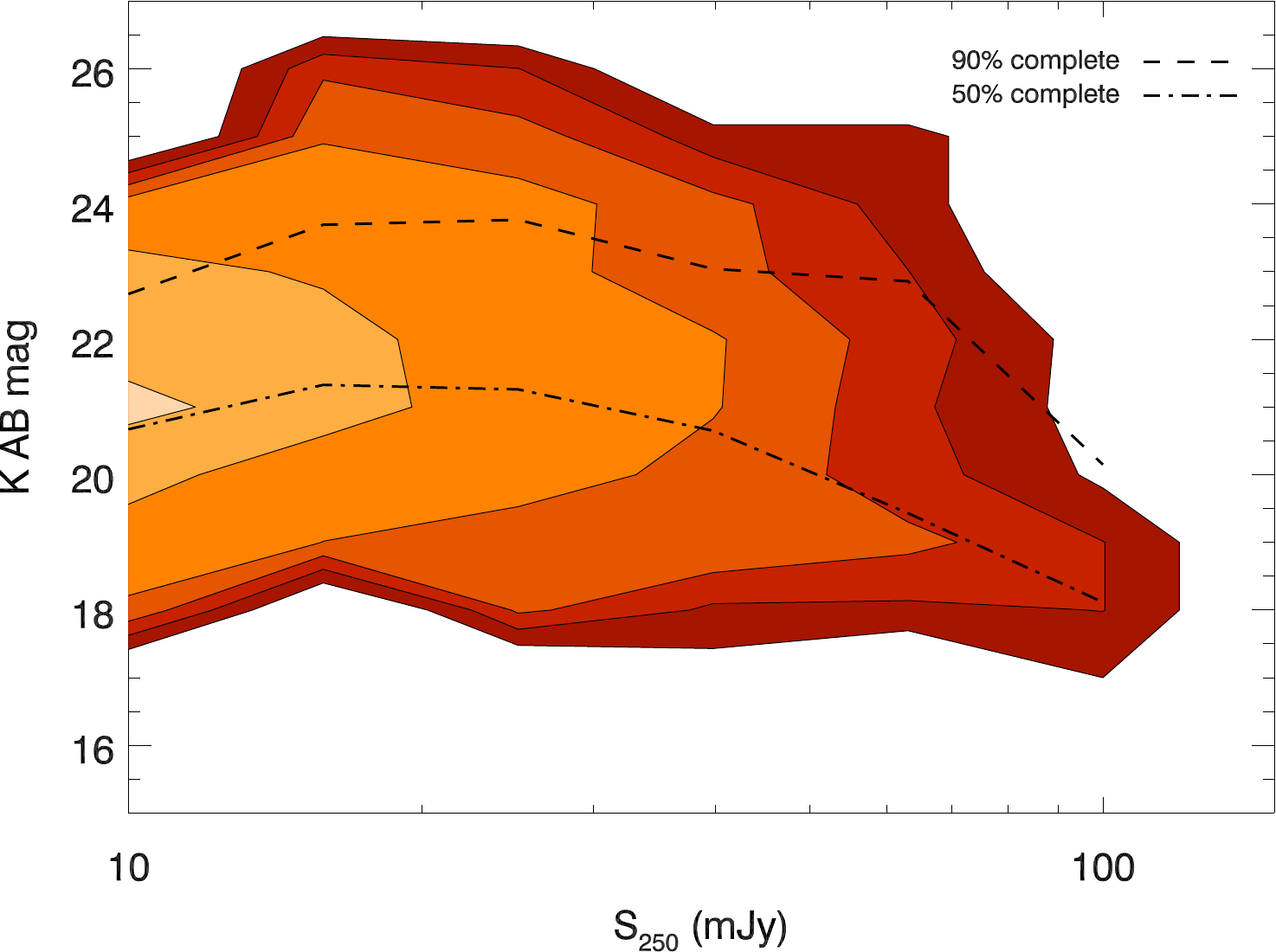}
\caption{Density of SPIRE sources as a function of 250$\,$\micron\ flux density and optical, $i$-band (top) and $K_{\rm s}$ (bottom) magnitudes.  The dashed line indicates the optical or NIR depth required to detect 90 per cent of the sample at a given 250$\,$\micron\ flux density, while the dot-dashed lines is the depth to detect 10 per cent.}\label{fig:compi}
\end{figure}

In Fig.~\ref{fig:compi} we show the number of sources as a function of 250$\,$\micron\ flux and $i$ or $K_{\rm s}$ band magnitude.  

\begin{table}
\begin{tabular}{llccc}
Band &  Units & 10\%   & 50\% & 90\% \\ \hline
UV(0.2\AA) & [AB] & 22.0 & 28.3 & 33.7\\
$R$ & [AB] &  18.6	& 22.5  & 25.2 \\
$I$ & [AB] &  18.1   & 21.5  & 23.6 \\
$K$ & [AB] &  17.2   & 19.5  & 20.8 \\
3.6$\,$\micron\ & [$\mu$Jy] & 380 &   90 & 30\\
24$\,$\micron\ & [$\mu$Jy] & 3000 & 880 & 220\\
70$\,$\micron\ & [mJy] & 42 & 13 & 4.7 \\
850$\,$\micron\ & [mJy] & 6.8 & 2.3 & 1.1 \\
21~cm & [$\mu$Jy] & 330 & 100 & 50 \\
\end{tabular}
\caption{Estimates of depth required to detect SPIRE galaxies at various other wavelengths.  The estimates are based on the mock catalogues of \protect\cite{Xu2003} cut to have $S_{\rm 250} > 30$mJy.  We tabulate the depth at which a given percentage of the catalogue would be detected.}
\label{tab:compdata}

\end{table}
 
\subsubsection{Available Ancillary data}
The survey fields are very well studied and it is outside the scope of this paper to provide a complete description of all the many ancillary data that are available in these fields.   A more detailed description of the ancillary data will be provided by Vaccari et al. (in prep.).   
Our intention is to homogenise and make publicly available all ancillary/complementary data in our final data release.

\subsubsection{Deliverable data products}

\begin{table*}
\begin{tabular}{|l|p{6cm}|p{6cm}|}\hline
Name & Description & Minimum Parameters\\\hline
SCAT & SPIRE Source Catalogues &Positions, Fluxes, errors, SNRs, etc.\\
SMAP & SPIRE Maps &Maps of flux, noise and coverage\\
PCAT & PACS Source Catalogues& Positions, Fluxes, errors, SNRs, etc.\\
PMAP & PACS maps &Maps of flux, noise and coverage\\
SPCAT & SPIRE/PACS band-merged catalogues &Positions, Fluxes, errors, SNRs, etc.\\
CLUS & Catalogues \& Maps for Clusters & As above for maps and catalogues\\
XID &  Cross identifications with selected homogenous catalogues at other wavelengths. &  Fluxes, errors, SNRs, positions, positional offsets \\\hline
\end{tabular}
\caption{Deliverable Data Products.}\label{tab:DataProducts}
\end{table*}

\noindent

Our intended data products are summarised in Table~\ref{tab:DataProducts}. The {\em Herschel} source catalogues from SPIRE and PACS data (SCAT and PCAT respectively) will consist of the usual independent lists where sources are
selected from data at one wavelength without reference to any other.  Associated with these catalogues will be validation analyses, including completeness, reliability and the information necessary to construct selection functions for standard scientific analysis. In addition these products will include fluxes estimated for sources from other catalogues (including sources from public  {\em Spitzer} catalogues).
Our first SCAT products are described in \citet{Smith2010} and our first PCAT products by Aussel et al. (in prep.).

The SPCAT product will include all {\em Herschel} bands. Upper limits will be listed for sources detected in
some {\em Herschel} bands but not others.

The XID product will include associations with a variety of large homogenous catalogues, including, but not necessarily limited to, public  {\em Spitzer} catalogues.  Our first XID products are described by \citet{Roseboom2010}.

Maps from SPIRE and PACS data (SMAP and PMAP respectively) will be suitable for extended source analysis, fluctuation analysis etc.
 Our first SMAP products are described by \citet{Levenson2010}.

\subsubsection{Other Data Products}
We expect to produce additional data products as an output of the pursuit of our science goals.  These
will include maps and catalogues of sources from data acquired at other facilities (optical, near-IR, radio etc.).  It will also include value-added products where observed data have been used to model other properties of the catalogued objects, such as photometric redshift, luminosity or spectral class.
It is impossible to define a complete list of such products at this stage.  We will make these available to the community on a best-efforts basis.
\subsubsection{Simulated data}
In order to plan our surveys and simulate our expectations we have compiled and homogenised mock catalogues from these and other models, which are publicly available via  \verb+hermes.sussex.ac.uk/+.  These and other simulations will be made available on a best-efforts basis through this site.

\subsubsection{Data Release Schedule}

\subsubsection*{Early Data Release: EDR}
Our first data release was proposed to be in time for the second open call for {\em Herschel} proposals (OT2).  This was before the Science Demonstration Phase (SDP) release rules were established and when OT2 was expected to be earlier.  In fact our SDP Early Data Release was made on 2010~July~1. This meant it was in time for OT1 (due on 2010~July~22).  This data release is described in \citet{Smith2010} and, as we proposed, it was restricted to SPIRE high signal-to-noise sources in order to be as reliable as possible.  It included maps from our Abell~2218 observation (\#1) and 250$\,$\micron\ catalogues limited at $S_{250}>100$~mJy for all our SDP fields (FLS\#40, GOODS-N \#14, Lockman-SWIRE \#28, Lockman-North \#19).

{\color{black} A second Early Data Release EDR2 was made on 2011~September~19 which included bright source catalogues similar to those for EDR but for the DR1 fields (see Table~\ref{tab:AORs}.)}

\subsubsection*{Data Release 1: DR1}
{\color{black} An extensive Data Release (DR1) of maps and catalogues will be made on 2012~March~27.} 
DR1 will include data from the SDP observations and all SPIRE observations completed by 2010~May~1 (A2219~\#7, MS0451.6-0305~\#3, ECDFS~\#15, XMM-LSS~\#36, EGS~HerMES~\#29, Groth Strip~\#17, Bo\"otes~\#37, ADFS~\#38, ELAIS~N1~HerMES~\#31).   All products will be accompanied by documentation in the form of papers in refereed journals.  

\subsubsection*{Data Release 2: DR2}
DR2 will occur at the end of the mission.  This will include all our deliverable data products and ancillary data in their final form.

\subsection{Archival Value and Data Access}
As our observations are in all the most well-studied survey fields, the legacy value is 
enormous.  We fully expect a rich data-base, leading to abundant science beyond the resources of 
our team.   In addition to any ESA data releases (\verb+herschel.esac.esa.int/+) our data will be released through the {\it Herschel} Database in Marseille, HeDaM (\verb+hedam.oamp.fr/HerMES+).
The information system design and its implementation are developed under the {\sc SItools} middleware interface provided by the CNES (\verb+vds.cnes.fr/sitools/+). The data (images and catalogues)  are accessible in various formats (fits files, VOTable, ascii) and accessible through Virtual Observatory Tools. Advanced searches, cross correlated data  and the corresponding images are also implemented, including visualization facilities like {\sc ALADIN} (\verb+http://aladin.u-strasbg.fr/+) and  {\sc TOPCAT} (\verb+http://www.star.bris.ac.uk/~mbt/topcat/+).

\section{Discussion and Conclusion}\label{sec:conc}
We have presented the Herschel Multi-Tiered Extra-galactic Survey (HerMES). This survey builds on the legacy of existing FIR and sub-mm surveys.  It will provide a census of star-formation activity over the wavelengths where the obscured star-formation peaks and over representive volumes (and thus environments) of the Universe at different epochs.  It is being carried out in some of the best studied extra-galactic fields on the sky, which is invaluable for the interpretation of the data both technically, by enabling accurate identifications and reducing the impact of confusion noise, and scientifically, by allowing exploration of the physical processes manifest at different wavelengths.  We have provided the description and rationale of the survey design. 
We also described the data products we plan 
to deliver and their schedule. 

Our first results from the Science Demonstration Phase data have fully demonstrated the promise of the full survey. We have quantified the confusion noise at SPIRE wavelengths \citep{Nguyen2010}, $5\sigma_{250}=29.0\pm1.5$~mJy, finding it to be very similar to what was anticipated.  This confusion is challenging to deal with \citep[e.g.][]{Brisbin2010} but we are exploring sophisticated techniques to deal with this \citep[e.g. through prior positional information, ][]{Roseboom2010} and using $P(D)$ analysis have already probed to $4\,{\rm mJy}$ and accounted for  64 per cent of the background at 250$\,$\micron\ \citep{Glenn2010}. It seems that previous phenomenological galaxy populations need revision \citep{Oliver2010, Glenn2010}  and we now anticipate that we will be able to catalogue over 100,000 galaxies with $>5\sigma$ detections at 250$\,$\micron.  The galaxies appear to be the luminous actively star-forming galaxies we expected \citep[e.g.][]{Elbaz2010} with a strongly evolving luminosity function \citep{Vaccari2010, Eales2010}.   Also, as expected, SPIRE probes a wide range of effective temperatures, including warm galaxies and those cooler galaxies typically seen by sub-mm surveys \citep{Hwang2010, Magdis2010, Chapman2010, Roseboom2011}.  A clue to the problems that the phenomenological models have may lie in the hints of the presence of cooler than expected dust in some galaxies \citep{Rowanrobinson2010,Schulz2010}.  We also see evidence for sources being magnified through gravitational lensing by foreground galaxies in the field \citep{Schulz2010, Wang2011, Conley2011}, and in targeted clusters. These  magnified galaxies provide a window to study intrinsically lower luminosity galaxies at higher redshifts.   We have identified strong clustering of SPIRE galaxies  \citep[e.g.][]{Cooray2010, Amblard2011}, indicating that these luminous systems lie in massive dark matter halos and implying they are the progenitors of galaxies in rich groups and clusters today, i.e. elliptical galaxies.

HerMES will constitute a lasting legacy to the community, providing an essential complement to multi-wavelength surveys in the same fields and  providing targets for follow-up using many facilities, e.g.  ALMA. The results are expected to provide an important benchmark for theoretical models of galaxy evolution for the foreseeable future.





\section*{Acknowledgements}
We acknowledge support from the UK Science and Technology Facilities Council [grant number ST/F002858/1] and [grant number ST/I000976/1]
HCSS / HSpot / HIPE are joint developments by the Herschel Science Ground 
Segment Consortium, consisting of ESA, the NASA Herschel Science Center, and the HIFI, PACS and 
SPIRE consortia.

SPIRE has been developed by a consortium of institutes led
by Cardiff Univ. (UK) and including Univ. Lethbridge (Canada);
NAOC (China); CEA, LAM (France); IFSI, Univ. Padua (Italy);
IAC (Spain); Stockholm Observatory (Sweden); Imperial College
London, RAL, UCL-MSSL, UKATC, Univ. Sussex (UK); Caltech,
JPL, NHSC, Univ. Colorado (USA). This development has been
supported by national funding agencies: CSA (Canada); NAOC
(China); CEA, CNES, CNRS (France); ASI (Italy); MCINN (Spain);
SNSB (Sweden); STFC, UKSA (UK); and NASA (USA).

\bibliography{Sebspaperfinal,hermes,hermesMNRAS409}


\appendix

\section{Detailed rational for definition of each survey region}\label{sec:fieldrational}

Our deepest tier, Level~1 (\#13), covers the GOODS-S region which is one of the two deepest  {\em Spitzer} fields \citep{goods}. 

The other GOODS field, GOODS-N, is covered by one of our Level~2 observations (\#14), though our observations are substantially wider.  The boundaries of our other Level~2 field, the Extended {\em Chandra} Deep Field South field (ECDFS, \#15), is defined by the deep FIDEL coverage \citep{fidel}.  

Our Extended Groth Strip (EGS) field at Level~3 (\#17) is also defined to match the FIDEL boundaries. The Lockman-East field at Level~3 (\#18, \#18B) covers  {\em Spitzer} guaranteed time program data (\#18) and the  {\em Spitzer} Legacy program of Egami et al. (\#18B). Those deep sets (\#13, 14, 15, 17 and 18) were all co-ordinated with the `PACS evolutionary Probe \citep[PEP,][]{pep} team.
The Lockman-North field at Level~3 (\#19, 20) covers the deep  {\em Spitzer} field defined e.g. in \citet{lock-n}.  

The UDS field at Level~4 (\#23) is defined by the  {\em Spitzer} SpUDS observations \citep{spuds} and we observe this field at Level~3 (\#21) with PACS.  The  {\em Spitzer} COSMOS field is observed in \#22 and \#22B, though our principal definition was the PEP observation of this field (discussed more in Section~\ref{sec:constraints}).  The VVDS field at Level~4 (\#24, 26) is not defined by  {\em Spitzer} observations but by the optical spectroscopic survey of \citet{vvds}.  

The Level~5 and 6 fields  CDFS~SWIRE, Lockman~SWIRE, XMM-LSS~SWIRE, ELAIS~N1~SWIRE, ELAIS~N2~SWIRE (\#27, 28, 34-36, 39 and 41) are defined by the SWIRE fields \citep{Lonsdale2003} -- those fields based in turn on the European Large Area {\em ISO} Survey, ELAIS, \citep{Oliver2000}; the XMM-LSS Survey \citep{xmmlss} and flanking the Chandra~Deep~Field~ South \citep{cdfs} and various Lockman Hole fields \citep{lockman}].  The Bo\"otes~NDWFS field at Level~6 (\#37) is defined by the  {\em Spitzer} Guaranteed time survey \citep{bootes}.  The FLS field Level~6 (\#40) is defined by the Extragalactic part of the  {\em Spitzer} First Look Survey \citep{fls} and is commonly referred to now as XFLS.   The {\em AKARI} deep field south (ADFS, \#38) is defined with reference to the  {\em Spitzer} \citep{adfs2a,adfs2}  and BLAST observations (but see Section~\ref{sec:constraints}).  The Level~5 observations in \#29, 30, 31, 32, 39B lie within or include other fields but the bounding regions are new (hence labelled `HerMES' or VIDEO) and have been planned with the expectation of subsequent follow-up with the SCUBA-2 Cosmology Legacy Survey \citep{s2cls}, the  {\em Spitzer} SERVS survey \citep{servs} and the VISTA-VIDEO survey \citep{video}.  The fields \#29, 32 and 39B were jointly defined in co-ordination with VISTA-VIDEO who fixed the final field location.

{\color{black}
\section{Modelling of SPIRE dithering patterns}\label{appendix:dither}
SPIRE maps are built by scanning an array of bolometers across the sky in a raster with long scan legs each separated by a short step, $\theta_{\rm max}$ (e.g. $\theta_{\rm max}= 348\arcsec$ for SPIRE `Large Map' mode).  The resulting hit-rate or coverage of detector 
readouts per sky bin is non-uniform (an effect which is exacerbated by dead or noisy bolometers). This non-uniform coverage can be improved by `dithering', i.e. repeating scan with offsets. We have modelled this to try and optimise the dithering pattern.

Since we are interested in point sources we can assume that the detector readouts will be combined with a point source filter \citep[e.g.][]{Smith2010}. The flux estimator for a source, $\hat{f}$ will be given by the 
$$\hat{f}=\frac{\sum_i{w_id_i/P_i}}{\sum_i{w_i}}$$ where $d_i$ is the readout of detector $i$, $P_i$ is the point source profile for the source at detector $i$ and $w_i$ is a weighting.  The optimal filter for isolated sources is  $w_i=P_i^2/\sigma_i^2$, where $\sigma_i$ is the noise of the detector $i$.  The variance in this estimator is
\begin{equation}
V=\sigma_{\hat{f}}^2=\frac{1}{\sum_i{w_i^2}}=\left( \sum_i{\frac{P_i^2}{\sigma_i^2}}\right)^{-1}.
\label{eqn:variance}\end{equation}  

We can consider the two scan directions independently so we need only model the coverage in one-dimension.  The sequential scan legs introduce a symmetry on the scale $\theta_{\rm max}$, so we project each bolometer position onto the range $0<\theta_i<\theta_{\rm max}$ in the cross-scan direction.
We then construct a one-dimensional variance profile $V(\theta)$ by analogy with equation~\ref{eqn:variance} setting $P_i$ the point spread function $P(\theta-\theta_i)$.   For simplicity we set $w_i=0$ for dead or noise bolometers and $\sigma_i=1$ otherwise and used Gaussian beams with {\sc FWHM}=18.15/25.15/36.3\arcsec\ for the three bands. 

We then defined a metric, $M$, to optimise dither patterns, on the understanding that we want to reduce the variation in variance.  When considering the dither pattern for one band in isolation we simply used the fractional variance of the variance 
$$M^2=\left\langle \sum{\left(\frac{V-\bar{V}}{\bar{V}}\right)^2} \right\rangle,$$
where the sum is over the profile.  As the SPIRE bolometers scan the sky simultaneously in all bands any dithering scheme would apply to all bands.  However, considering three bands simultaneously there is no obvious metric (unless we considered sources of a particular colour) we did define an arbitrary metric $M^2=M_{\rm PSW}^2+M_{\rm PMW}^2+M_{\rm PLW}^2$ but have restricted this discussion to the single bands independently.

The aim is to choose a dither pattern that minimises $M$. If  $N$ independent scan maps with $N-1$ dither positions are available then the dither pattern is defined by $N-1$ offsets ${\bf \Delta\theta}$.  We adopted four approaches: (a) optimisation by brute-force search through $N-1$ dimensional space (only attempted up to $N=4$) (b) sequential optimisation where we chose the best $\Delta\theta_i$ for each additional dither, $i$, given the ${\bf \Delta\theta}$ found for the  previous $i-1$ dithers (c) equal spacing $\Delta\theta_1=\Delta\theta_2=\ldots=\theta_{\rm max}/N$ (d) random spacing with $\Delta\theta_i$ uniformly selected from $0<\Delta\theta_i<\theta_{\rm max}$.  

For low values of $N\le4$ where both were calculated we found that the brute-force optimisation (a) agreed reasonably well with the sequential optimisation (b).  We found that the equal spacing (c) performed similarly to the sequential optimisation at low $N$ and typically better at high $N\ga10)$ except at specific $N$  (e.g. $N=15$ for PSW and $\theta_{\rm max} = 348\arcsec$) when the projected bolometer spacing were in phase.  Random offsets (d) were invariably worst.  The raw variation with no dithers ($N=1$) was 12, 15, 10 per cent for PSW, PMW and PLW respectively, this declined rapidly to about 3 per cent by $N=3$ and was $<1$ per cent for $N>16$.

A penalty for dithering with these large steps is that the  ramp down in coverage at the edges of the map is more gradual, i.e. less area at the full coverage with more area at low coverage.  When designing offsets in both scan directions we chose pairs of offsets tracing a square to reduce the impact of this ramp-down and this strategy is included in the SPIRE Observers' Manual.}
\end{document}